\documentclass[journal=jacsat,manuscript=article]{achemso}

\usepackage[version=3]{mhchem} % Formula subscripts using \ce{}

\author{Fu-Chen Hsiao}
\affiliation{Department of Electrical and Computer Engineering, University of Illinois at Urbana-Champaign, Urbana, Illinois 61801, United States}
\email{fhsiao3@illinois.edu}

\author{Arnab Hazari}
\affiliation{Department of Electrical Engineering and Computer Science, University of Michigan, 1301 Beal Avenue, Ann Arbor, Michigan 48109-2122, United States}

\author{Pallab Bhattacharya}
\affiliation{Department of Electrical Engineering and Computer Science, University of Michigan, 1301 Beal Avenue, Ann Arbor, Michigan 48109-2122, United States}

\author{Yia-Chung Chang}
\affiliation{Research Center for Applied Sciences, Academia Sinica, Taipei 11529, Taiwan}

\author{John M. Dallesasse}
\affiliation{Department of Electrical and Computer Engineering, University of Illinois at Urbana-Champaign, Urbana, Illinois 61801, United States}
\title{Modeling Photocurrent Spectra of In$_{0.91}$Ga$_{0.09}$N/In$_{0.4}$Ga$_{0.6}$N Disk-in-Wire Photodiode on Silicon for $1.3$ $\mu$m $-$ $1.55$ $\mu$m Operation}

\abbreviations{IR,NMR,UV}

\begin{document}
%%%%%%%%%%%%%%%%%%%%%%%%%%%%%%%%%%%%%%%%%%%%%%%%%%%%%%%%%%%%%%%%%%%%%
%% The manuscript does not need to include \maketitle, which is
%% executed automatically.  The document should begin with an
%% abstract, if appropriate.  If one is given and should not be, the
%% contents will be gobbled.
%%%%%%%%%%%%%%%%%%%%%%%%%%%%%%%%%%%%%%%%%%%%%%%%%%%%%%%%%%%%%%%%%%%%%
\begin{abstract}
 This work reports comprehensive theoretical modeling of photocurrent spectra generated by an In$_{0.91}$Ga$_{0.09}$N/In$_{0.4}$Ga$_{0.6}$N disk-in-wire photodiode. The strain distribution is calculated by valence-force-field (VFF) model, while a realistic band structure of the InN/InGaN heterostructure is incorporated using an eight-band effective bond-orbital model (EBOM) with spin-orbit coupling neglected. The electrostatic potential is obtained from self-consistent calculation employing the non-equilibrium Green's function (NEGF) method. With the strain distribution and band profile determined, a multi-band transfer-matrix method (TMM) is used to calculate the tunneling coefficients of optically-pumped carriers in the absorbing region. The photocurrent spectra contributed by both single-photon absorption (SPA) and two-photon absorption (TPA) are calculated. The absorption coefficient is weighted by the carrier tunneling rate and the photon density-of-state (DOS) in the optical cavity formed in the nanowire region to produce the photocurrent. The calculated photocurrent spectra is in good agreement with experimental data, while physical mechanisms for the observed prominent peaks are identified and investigated.
\end{abstract}

\section{Introduction}
Integration of III-V devices on silicon is a well-known technique for overcoming the low efficiency of photon-absorbing and emitting process in Si-based photonic devices\cite{INTRO_1,INTRO_2,INTRO_3}. The direct epitaxial growth of III-V materials and heterostructure on silicon appears to be the most straightforward method to achieve III-V on silicon integration \cite{INTRO_4,INTRO_5,INTRO_6}. However, due to the high density of threading dislocations, mismatch of thermal expansion coefficients, and formation of antiphase domains, the direct epitaxial growth method encounters many problems that have not been properly resolved.

Instead of planar structures, growing nanowire structures consisting of III-nitride (III-N) semiconductors in WZ phases provides a very different means to deal with the problem of III-V/Si integration \cite{INTRO_7,INTRO_8,INTRO_9}. Due to efficient lateral-strain relaxation at the nanowire-silicon interface, nanowire heterostructures show efficient reduction of polarization fields\cite{INTRO_10} and achieve smaller radiative recombination time than a system with quantum well (QW) structures \cite{INTRO_7}. From the simulation perspective, the modeling for high-indium-content III-N based device in the WZ phase presents multiple challenges. The ambiguity introduced by the ordering of the operators for kinetic energy may lead to spurious solutions in the $k \cdot p$ model\cite{KPsp1,KPsp2}, especially in the large wavevector region for high-indium-content devices due to the uncertainty of band parameters in InN. This indicates that a band structure model which is capable of producing accurate band structure in the full Brillouin Zone (BZ) is needed. Besides that, hexagonal structures present much lower symmetry ($C_{3v}$) than the cubic phase ($T_d$), which requires extra fitting parameters for the band structure and effective-masses. As a result, a semi-empirical model which is capable of producing full-zone band structure and effective-masses close to experimental values near the zone center is needed. Furthermore, since the optical transition does not necessarily happen near the BZ center for high-indium-content devices, the optical matrix elements which includes wavevector dependence are required as well. Recently, a novel monolithic optical interconnect on $(001)$Si substrate using an InGaN disk-in-wire array, which consists of an edge emitting diode laser and guided wave photodiode, has been demonstrated experimentally \cite{INTRO_11}. To the best of our knowledge, due to the complexity of the geometry, ultra-high indium content active region, and the requirement of a realistic band structure calculation within a large portion of first BZ, a systematic study of the photon absorption process as well as the carrier tunneling behaviors is not yet available.

In this work, we provide a comprehensive theoretical analysis for the photocurrent spectra of high-indium-content InN/InGaN disk-in-wire photodiodes with a guided wave structure base on the eight-band effective bond-orbital model (EBOM) \cite{EBOM1,EBOM2,EBOM3}. The theoretical model takes into account the strain distribution, polarization field effect (piezoelectric and spontaneous), and realistic band structure of InN/InGaN with suitable boundary conditions for modeling quantum carrier transport. The photon density-of-states (DOS) which is sharply peaked at the resonance modes of the InN/InGaN waveguide structure is also incorporated in our model by using the rigorous couple-wave analysis (RCWA) \cite{RCWA}. These sharp peaks are fitted with Lorentzian functions centered at cut-off energies with resonance bandwidth. The simulation provides detailed descriptions on the corresponding optical transition process for each of the peaks on the photocurrent spectra. The calculated photocurrent spectra which includes both one-photon and two-photon absorption processes is presented and compared with the measured data.
\section{Strain Distribution in In$_x$Ga$_{1-x}$N Disk-in-Wire Superlattice Structure}
\begin{figure}[H]
		\centering\includegraphics[width=1\columnwidth]{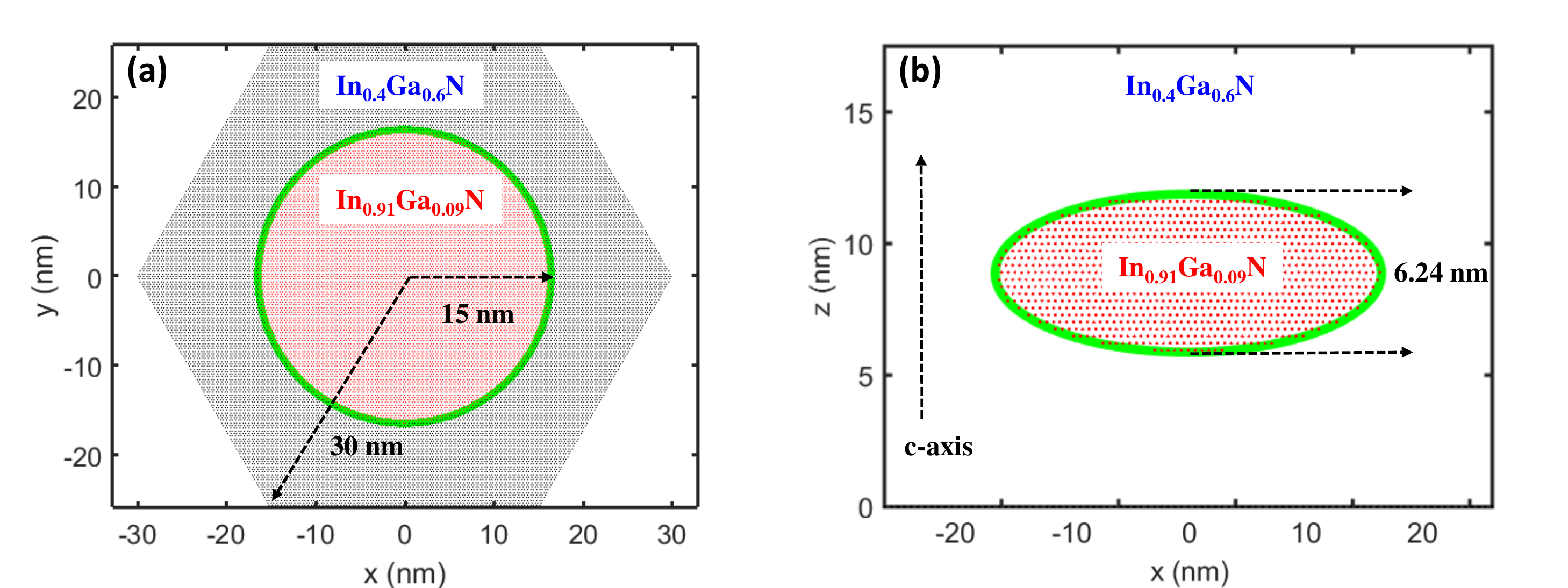}
	\caption{(a) Schematic diagram of a cross-section in the $x-y$ plane for In$_x$Ga$_{1-x}$N disk buried in In$_{0.4}$Ga$_{0.6}$N structure. (b) The side view of the same structure.}
\label{fig_4_1}
\end{figure}

The detailed description of the InN/InGaN disk-in-wire structure of interest can be found in Ref. \citenum{INTRO_11}. Here, we model it by using a superlattice structure along the c-axis, in which each unit cell contains an  In$_x$Ga$_{1-x}$N quantum disk (QD) with lateral diameter of $29.98$ nm and height of $6.24$ nm, embedded in a single In$_{0.4}$Ga$_{0.6}$N wire as illustrated in Fig. (\ref{fig_4_1}). The wire has a hexagonal cross-section with side length of $30$ nm and height of $17.75$ nm. The geometric parameters are obtained by examining the Transmission Electron Microscopy with High Angle Annular Dark Field (TEM-HAADF) and Energy Dispersive x-ray Spectroscopy (EDS) images in Ref \citenum{INTRO_11}. To calculate the strain distribution, we adopt the valence-force field (VFF) model with periodic boundary condition along the z-axis and free boundary condition on the boundary in the $x-y$ plane. Since the VFF model can give a strain tensor at the atomistic level and retain the correct point symmetry of the system, while avoiding potential failure at the interface \cite{VFF_1,VFF_2,VFF_3}, it is chosen in favor of the continuum mechanical model for this work.  All the parameters for III-N materials used in VFF model are adopted from Ref. \citenum{VFF_4}, where linear interpolation has been used for the intermediate In$_x$Ga$_{1-x}$N alloys. Given the local strain tensor, the strain-induced deformation potential is then incorporated in the EBOM Hamiltonian according to Bir-Pikus theory \cite{STRAIN_1}. The strain field influences the electronic states of a InN/InGaN QD via the strain-induced deformation potential and the piezoelectric polarizations. In addition to piezoelectric polarization, due to high electronegativity of nitrogen atom, the spontaneous (pyroelectric) polarization also exists along the c-axis of the WZ crystal.

\begin{figure}[H]
		\centering\includegraphics[width=1.\columnwidth]{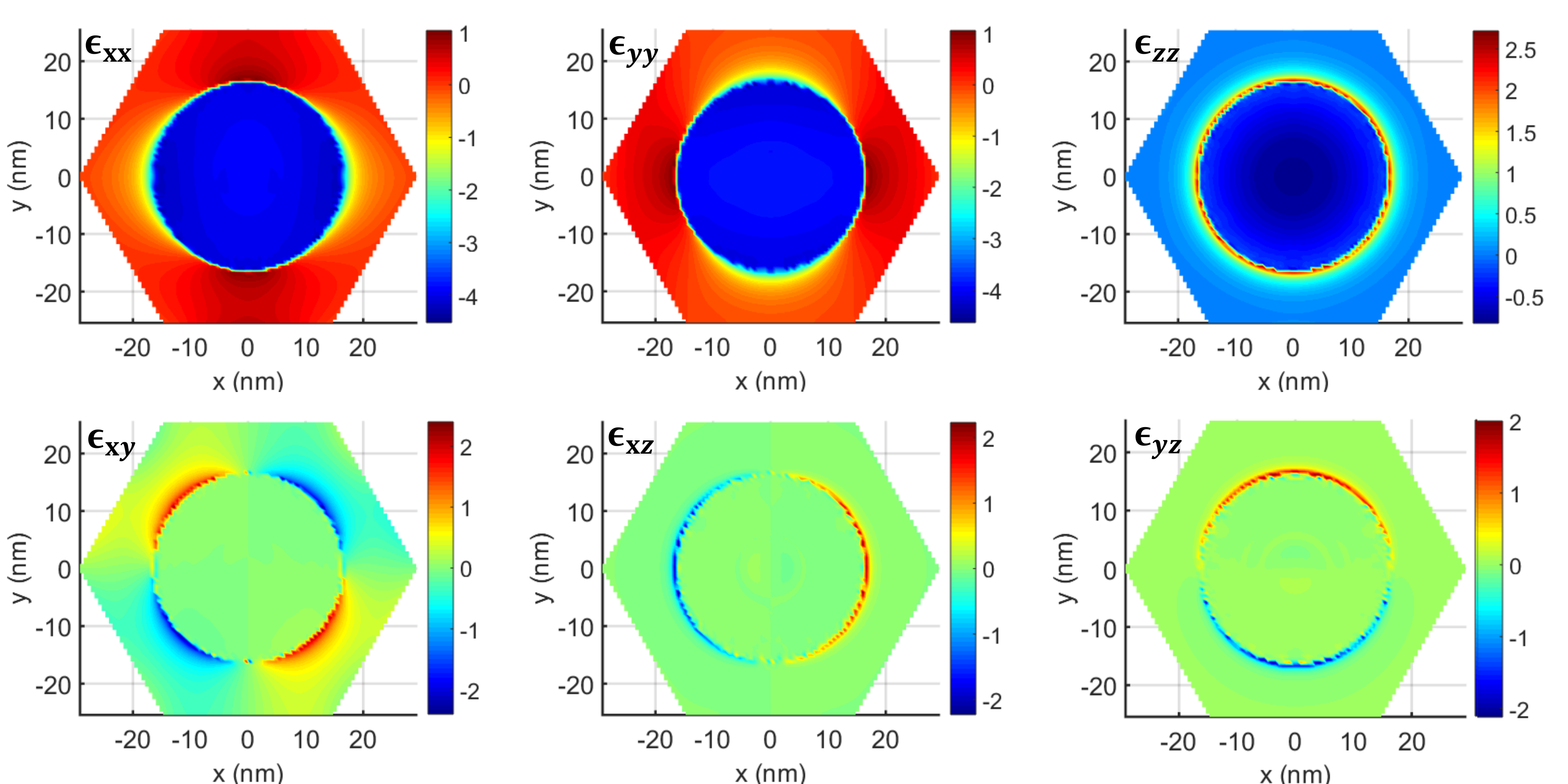}
	\caption{Calculated strain distribution in an In$_{0.91}$Ga$_{0.09}$N disk with $6.24$ nm height in the
plane passing through the center of the disk ($z = 8.64$ nm). The unit are in percentage.}
\label{fig_4_2}
\end{figure}

\begin{figure}[H]
		\centering\includegraphics[width=1.\columnwidth]{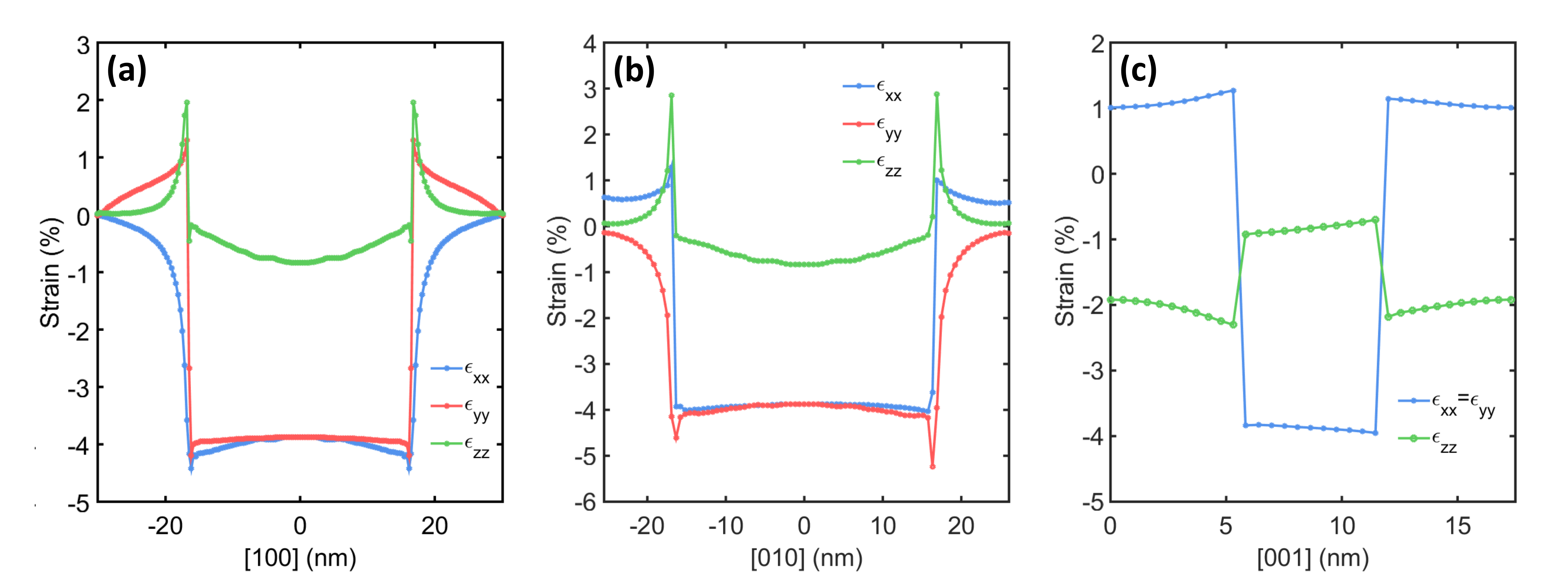}
	\caption{Calculated strain distribution in an In$_{0.91}$Ga$_{0.09}$N dot with $6.24$ nm height along a line passing through the center of the disk and parallel to (a) x axis, (b) y axis, and (c) z-axis.}
\label{fig_4_3}
\end{figure}

\begin{figure}[H]
		\centering\includegraphics[width=1.\columnwidth]{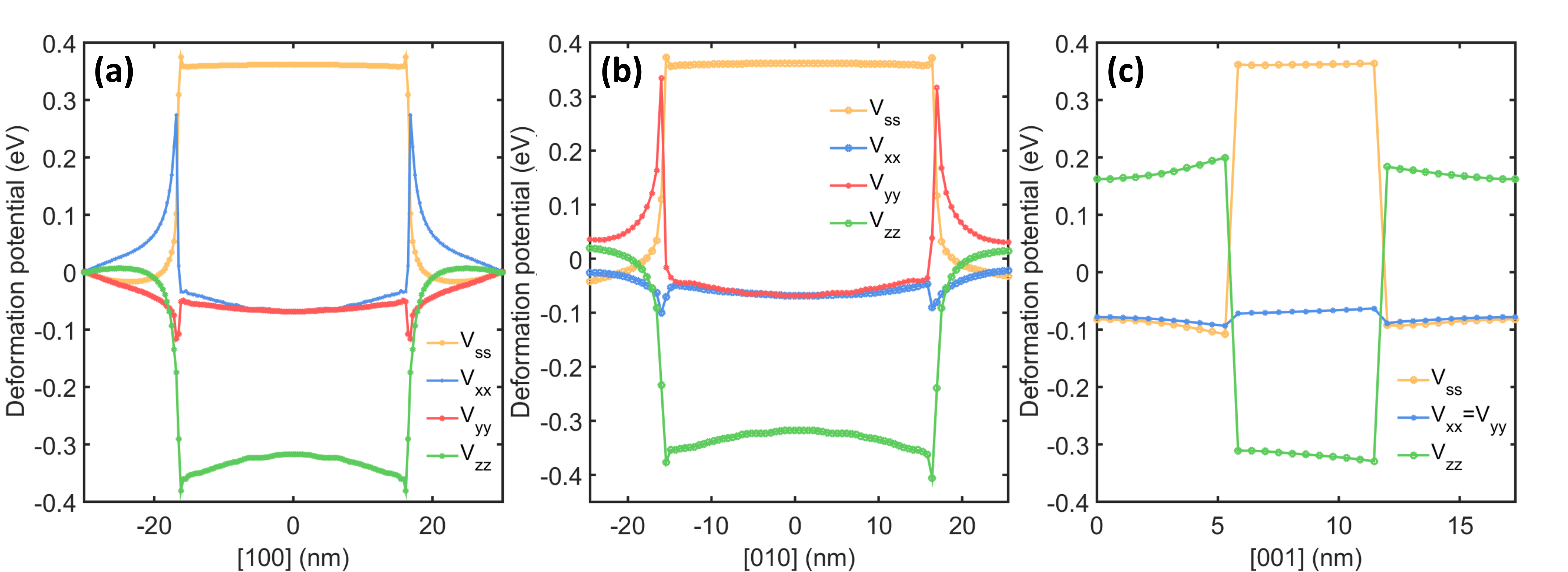}
	\caption{Diagonal elements of the deformation potential for In$_{0.91}$Ga$_{0.09}$N disk with $6.24$ nm height along a line passing through the center of the disk and parallel to (a) x axis, (b) y axis, and (c) z-axis.}
\label{fig_4_4}
\end{figure}

\begin{figure}[H]
		\centering\includegraphics[width=1\columnwidth]{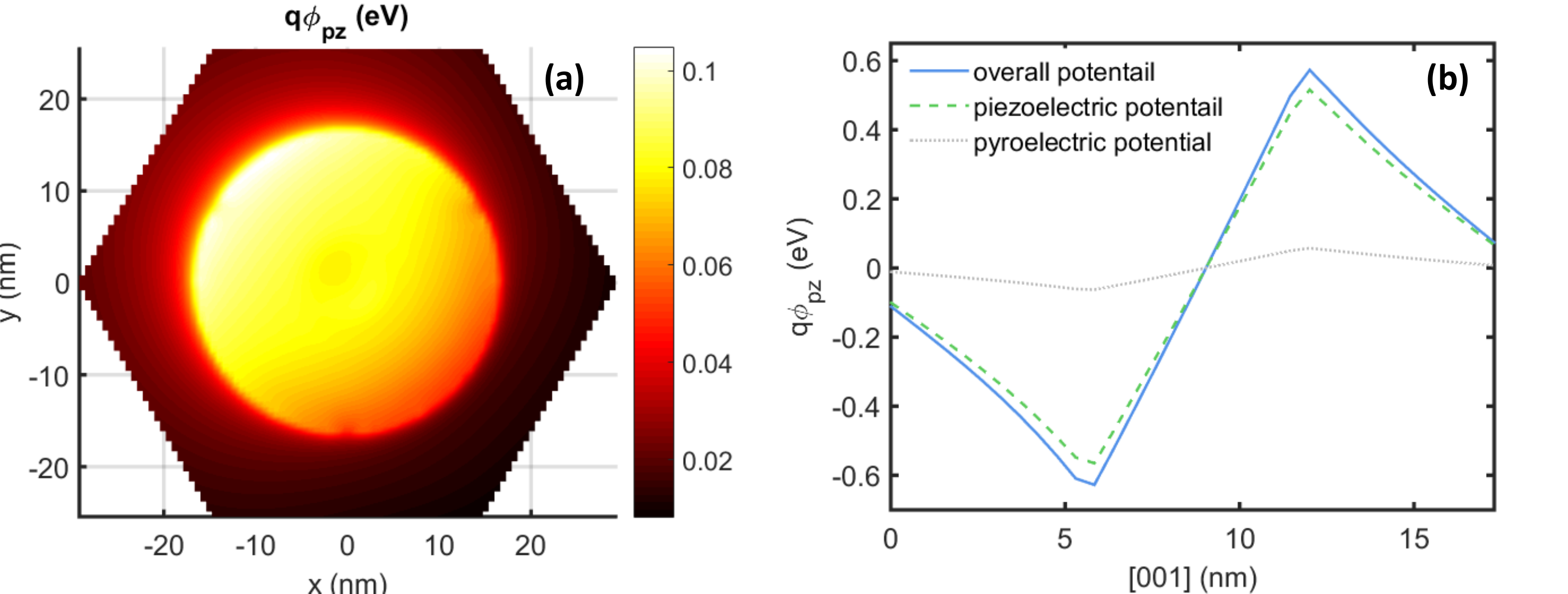}
	\caption{Calculated built-in electric potential in an In$_{0.91}$Ga$_{0.09}$N disk with $6.24$ nm height. (a) The overall electric potential map in the plane of $z=8.64$ nm. (b) The linescan of the built-in potential along the [$0001$] direction through the disk's center. The solid line shows the overall potential, and the dashed line and the dotted line show the components induced by the piezoelectric polarization and the pyroelecric polarization.}
\label{fig_4_5}
\end{figure}

The strain distribution in In$_{0.91}$Ga$_{0.09}$N disk-in-wire structure in the $x-y$ plane and along $z$-direction are shown in Fig. (\ref{fig_4_2}) and Fig. (\ref{fig_4_3}), respectively. Figure (\ref{fig_4_4}) shows the diagonal elements of the strain Hamiltonian along [100], [010], and [001] directions. Here $V_{ss}=a_2(\epsilon_{xx}+\epsilon_{yy})+a_1\epsilon_{zz}$, $V_{xx}=l_1\epsilon_{xx}+m_1\epsilon_{yy}+m_2\epsilon_{zz}$, $V_{yy}=m_1\epsilon_{xx}+l_1\epsilon_{yy}+m_2\epsilon_{zz}$, and $V_{zz}=m_3(\epsilon_{xx}+\epsilon_{yy})+l_2\epsilon_{zz}$. We see that the hydrostatic strain is approximately uniform throughout the disk, which justifies the divide-and-conquer approach to be discussed in Sec. (3). In addition, the deformation potential profiles show that the strain effects cause an extra contribution to the energy splitting between the $|s\rangle$ state and $|z\rangle$ state, which adds on the splitting between the conduction band and crystal-field split band. Figure (\ref{fig_4_5}) shows the potential energy in $x-y$ plane (Fig. (\ref{fig_4_5}.a)) and the decomposition of the overall potential energy into a piezoelectric and a pyroelectric components (Fig. (\ref{fig_4_5}.b)), which indicates that the built-in electric potential is mainly contributed by piezoelectric effects. In this case, the piezoelectric potential is about nine times larger than the pyroelectric potential. The modifications to the confinement potential by piezoelectric and pyroelectric effects are of the same order of magnitude as the material band offsets.

\section{Electronic States of Single In$_x$Ga$_{1-x}$N Disk-in-Wire Structure}
Here we describe how to calculate the electronic states of  a single In$_x$Ga$_{1-x}$N disk-in-wire structure. The model presented in this section will be used to determine the geometric parameters and indium mole fraction in the disk region by fitting the calculated inter-band transition energy with the dominant peak in the photoluminescence (PL) spectrum measured experimentally. The electronic states of QD depend significantly on the strain distribution in III-N system. Using the strain distribution calculated via VFF model as described in the previous section and the deformation potential, we can evaluate the electronic states by EBOM.

Since the lateral size of the disk region is fairly large ($\sim 30$ nm), the quantum confinement effect is more significant along the x-axis than in the $x-y$ plane. If we first neglect the lateral confinement due to finite size, the system can be described by a zero-th order EBOM Hamiltonian, $H^{(0)}$ suitable for describing  a quantum well (QW) along the c-axis with the strain-induced potential $V(x,y,z)$ in each layer replaced by its average value $V_0$ in the $x-y$ plane. We then include the effect of lateral confinement and the difference in strain-induced potential $\Delta V(x,y,z)=V(x,y,z)-V_0$  as a correction term $H^{(1)}$. The total Hamiltonian then reads
\begin{equation}
 H_{tot}=H^{(0)}+H^{(1)}.
 \label{eq_2_1}
\end{equation}
An eight-band EBOM for WZ crystal\cite{EBOM3} is applied to calculate the eigenstates of $H^{(0)}$. All the EBOM parameters used in this work are adopted from Ref \citenum{EBOM3} for the room-temperature case. The interactions between any two bond-orbitals located in the same material are taken to be the same as those for the bulk. The interaction parameters between two bond-orbitals located in two different materials (i.e. across the heterojunciton) are obtained by using the average of the corresponding matrix elements in the two participating bulk materials. The eigenstates of the $H^{(0)}$ matrix for the QW described in EBOM can be solved efficiently with a band-matrix diagonalization program. The resulting eigenstates are denoted by
$|n,{\bf k} \rangle$, where ${\bf k}$ is the wave vector in the $x-y$ plane and $n$ is the label of QW subbands.

To facilitate the calculation, the eigenfunctions of $ H_{tot}$ are expanded in terms of the eigenfunctions of $H^{(0)}$, and the lateral confinement effect is described by the truncated-crystal approximation \cite{BZ_CUT_2}. In this approximation the EBOM envelope function of a QW state in the lateral direction is approximated by a plane wave $e^{i{\bf k}\cdot {\bf r}}$ subject to suitable boundary conditions. Using the Jacobi-Anger  expansion in cylindrical coordinates, we have
\begin{equation}
e^{i{\bf k}\cdot {\bf r}}=\sum_m (i)^m J_m(k\rho)e^{im(\phi-\phi_k)}.
\end{equation}
Since the interface between the nanowire and ambient material has a large band offset, it is reasonable to choose a boundary condition such that all lateral wave functions vanish on the interface, which is approximated by a cylindrical surface with radius $r_0$. With this hard-wall boundary condition, the allowed values of $k$ are $\alpha^m_\nu =x_\nu^{m}/r_0$, where $x_\nu^{m}$ are zeros of the Bessel function $J_m(x)$. Thus, the basis states to use for the expansion for eigenstates of $H_{tot}$ are described by $|n,m,{\nu}\rangle$ with
\begin{equation}
\langle \rho, \phi |n,m,{\nu}\rangle = C^m_{\nu} J_m(\alpha \rho)e^{im\phi},
\end{equation}
where $C^m_{\nu}=1/\sqrt{\pi r^2_0 J^2_{m+1}(x^m_\nu)}$ is the normalization constant. Each basis function can be related back to the plane waves via the relation $J_m(\alpha\rho)e^{im\phi}=(-i)^m\int \frac{d\phi_k}{2\pi} e^{i{\bf k}\cdot {\bf r}}e^{im\phi_k}$. Hence, the matrix elements of $H^{(0)}$ in this basis reads
\begin{equation}
\begin{aligned}
 & \langle n',m',{\nu'}|H^{(0)}|n,m,\nu \rangle = \sum_{\bf k} \langle n',m',{\nu'}|{\bf k}\rangle \langle {\bf k} | H^{(0)}|{\bf k} \rangle \langle {\bf k} |n,m,\nu \rangle \\
                                            & = \int kdk \int d\phi_k  E^{ebom}_n(k,\phi_k)\delta_{n',n}I^{m',m}_{\nu,\nu'} \frac {J_{m'}(kr_0)J_m(kr_0)}{[k^2-(\alpha^{m'}_{\nu'})^2][k^2-(\alpha^{m}_{\nu})^2]}e^{i(m-m')\phi_k},
\end{aligned}
\end{equation}
where $E^{ebom}_n({\bf k})$ is the QW energy of subband $n$ at lateral momentum ${\bf k}$ obtained by EBOM and
\begin{equation}
I^{m',m}_{\nu,\nu'}=\frac 1 \i (-1)^{\nu'+\nu+m} (i)^{m'+m} \alpha^{m'}_{\nu'}\alpha^m_\nu.
\end{equation}
For the  $H^{(1)}$  term, which consists of only the potential part, the matrix elements can be written as
\begin{equation}
\langle n',m',{\nu'}|H^{(1)}|n,m,\nu \rangle =  \int f_{n',m',\nu'}^{*}(z)f_{n,m,\nu}(z)\Delta V^{m'm}_{\nu',\nu}(z) dz,
\label{eq_2_24}
\end{equation}
where $f_{n,m,\nu}(z)$ denotes the envelope function of QW along $z$ axis, which is related to the eigenvectors of QW for the $n$-th subband at $k=\alpha^m_\nu$ obtained from the eight-band EBOM and averaged over angle. We have used the root-mean-square value of bond-orbital coefficients on each site to get a smooth envelope function.
\begin{equation}
\Delta V^{m'm}_{\nu',\nu}(z)= \int_{0}^{2\pi} \int_{0}^{r_0} C_{\nu'}^{m'} C_{\nu}^{m}J_{m'}(\alpha_{\nu'}^{m'}\rho)J_{m}(\alpha_{\nu}^{m}\rho)\Delta V(\rho,\phi,z)\rho d\rho d\phi.
\end{equation}
In this manner, the electronic states in In$_x$Ga$_{1-x}$N disk-in-wire can be solved efficiently in the subspace of the full Hamiltonian defined by the selected subband states obtained from EBOM.
\section{Determination of Indium Mole Fraction in Disk Region}

\begin{figure}[H]
		\centering\includegraphics[width=1\columnwidth]{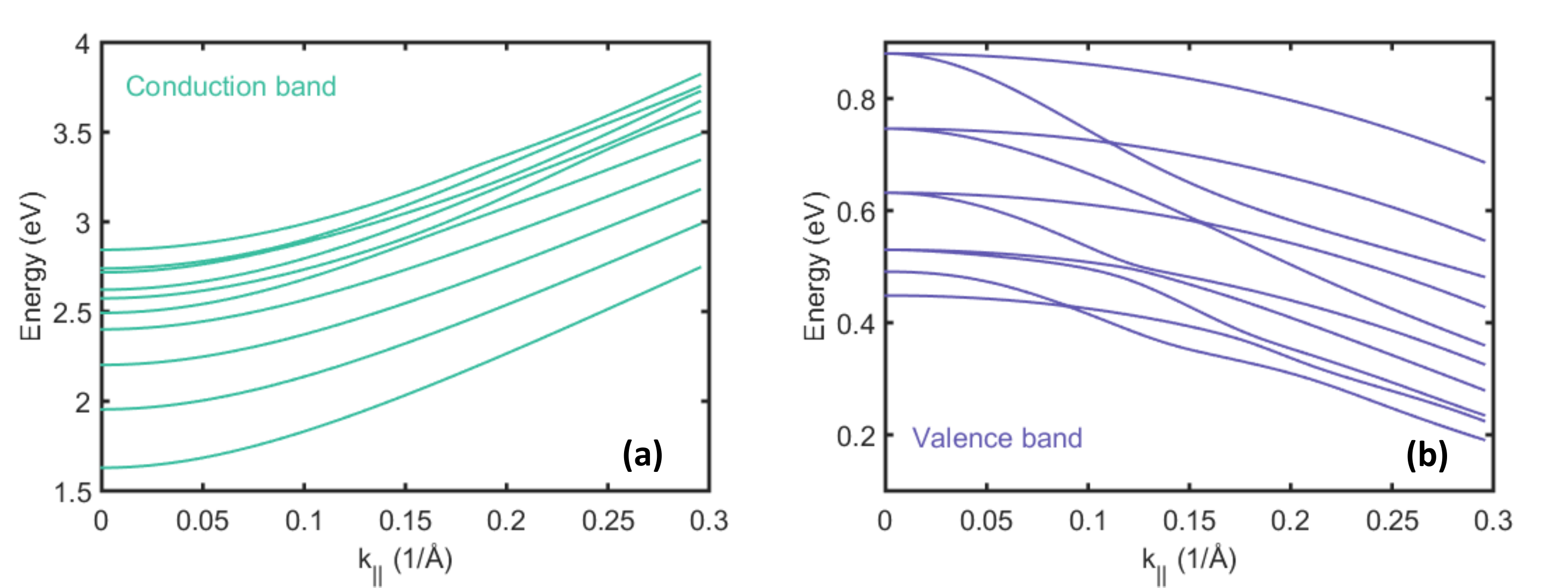}
	\caption{Subband structures of a $6.24$ nm width In$_{0.91}$Ga$_{0.09}$N QW for an electron in (a) conduction band and (b) valence band.}
\label{fig_4_6}
\end{figure}

\begin{figure}[H]
		\centering\includegraphics[width=1\columnwidth]{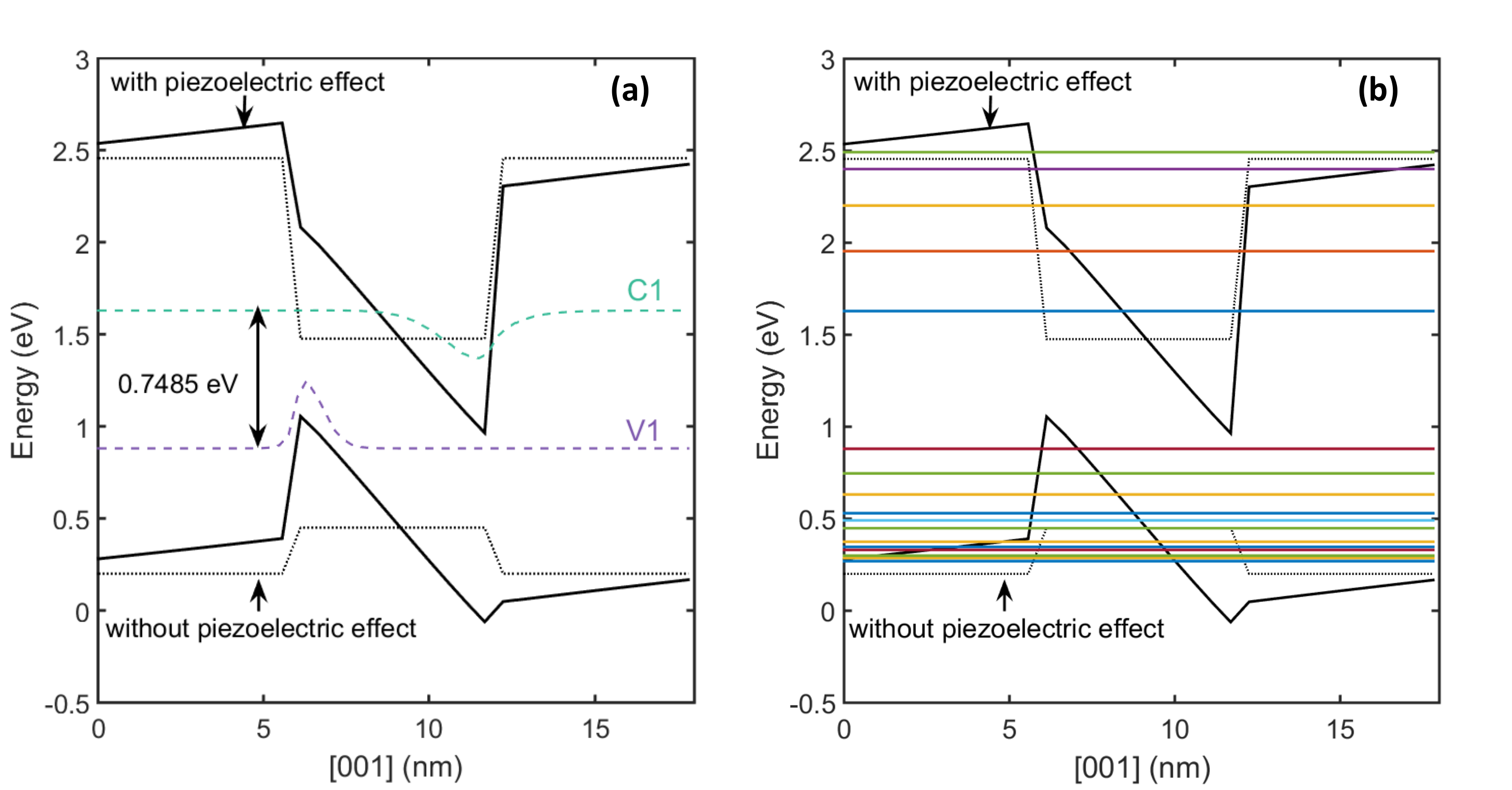}
	\caption{The in-plane averaged band profile for an In$_{0.91}$Ga$_{0.09}$N quantum well structure with $6.24$ nm width with (solid line) and without (dotted line) piezoelectric effect. (a) The corresponding ground state in conduction and valence band (dashed lines). (b) The bottom $5$ energy levels in the conduction band and top $20$ energy levels in the valence band (colored lines).}
\label{fig_4_7}
\end{figure}

In this section, we perform the calculation of electronic state of single In$_{x}$Ga$_{1-x}$N/In$_{0.4}$Ga$_{0.6}$N disk-in-wire structures, which is presented in Sec. (3), in order to determine the indium mole fraction at disk region by fitting the energy of the dominant PL peak. Figure (\ref{fig_4_6}) shows the subband structures of a $6.24$ nm-wide In$_{0.91}$Ga$_{0.09}$N disk in the lateral direction. Note that the top four valence bands are doubly degenerate at the zone center which corresponds to the heavy-hole and light-hole bands, while the single-fold states correspond to the split bands due to the crystal field. The crystal-field split bands are pushed down in energy due to the deformation potential contributed from the strain. Figure (\ref{fig_4_7}) shows the band profile in the growth direction as well as the calculated electronic states in an In$_{0.91}$Ga$_{0.09}$N QW structure. The dominant PL peak measured at $300$ K occurs at $0.7485$ eV ($z$-direction) $+ 0.0075$ eV (lateral direction) $=0.7560$ eV and the best-fit  indium composition inside the active region of the disk is $\sim91$\%.

\section{Carrier Leaking Rate in Quantum Well System: Multi-Band Transfer-Matrix Method}

In order to study the photocurrent spectra, we calculate the leaking rate of a carrier in the full epitaxial structure of the  device. There are multiple  theoretical methods for evaluating the carrier tunneling rate, including the phase-shift method \cite{ST_1}, the complex-energy method \cite{ST_2}, and the stabilization method \cite{ST_3,ST_4}. The main drawback for both the phase-shift method and the complex-energy method is the requirement of accurate description of wavefunctions. In contrast, the stabilization method requires only the eigenvalues of the system Hamiltonian as functions of an external parameter, such as the size of simulation domain. It can calculate the electron/hole leaking rate (or carrier lifetime) given the actual coupling strength between states, and the width of the generated DOS averaged by a scaling parameter. However, it is impractical when the device has a complicated structures and possesses multiple tunneling channels. Besides that, the stabilization method cannot really incorporate the continuum-to-continuum state transitions. As a result, we adopt an approach which is based on transfer-matrix method (TMM) to calculate the reflection coefficient spectra of the InN/InGaN disk-in-wire photodiode, and relate that to the leaking rate of bound or quasi-bound states in the disk region. Again, the band structure information is incorporated via an eight-band EBOM, while suitable boundary conditions are applied at the contact region (or the boundary of the simulation domain).

A numerically stable version of multi-band TMM base on EBOM for the devices composed of ZB semiconductor can be found in Ref \citenum{EBOM_TP}, which can be easily extended into the system with WZ materials. Here,  we set up the transfer-matrix equations for WZ material, and solve the resulting linear equation by using a band matrix solver. The reflection coefficients obtained from the TMM can be used to determine the leaking rate for the bound or quasi-bound states in the active region, which quantitatively describe the carrier tunneling in the active region through the barrier. The leaking rate so determined is utilized in the calculation of the photocurrent strength of InN/InGaN disk-in-wire photodiode.

\begin{figure}[H]
		\centering\includegraphics[width=1.\columnwidth]{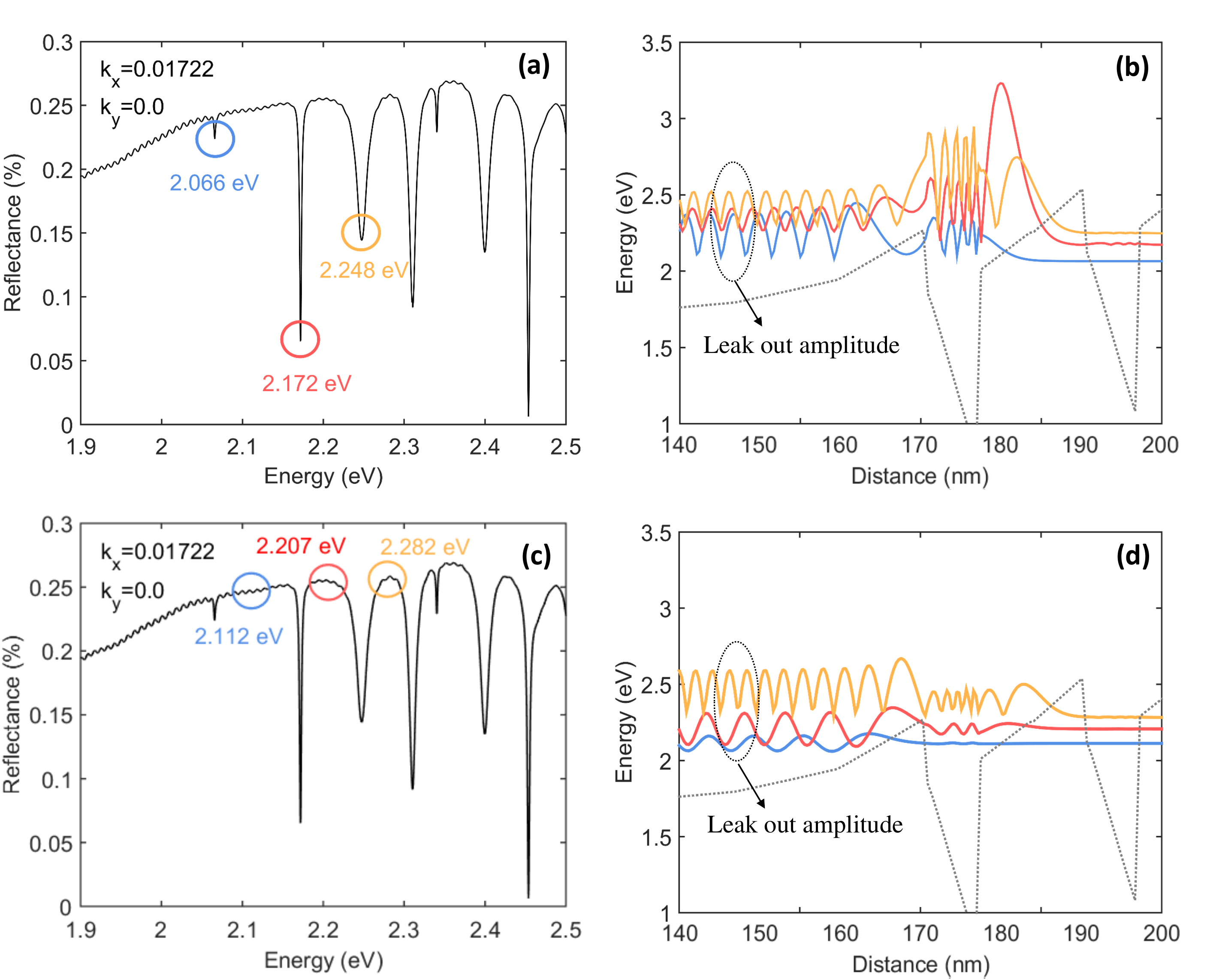}
	\caption{Reflection spectra and wave functions of quasi-bound states for InN/InGaN multi-QW structure calculated by EBOM-based TMM. (a)Reflection spectra with positions of quasi-bound states indicated by circles. (b) Wave functions of quasi-bound states with energies positions indicated by circles in (a). (c) Reflection spectra with a few selected non-resonant energy positions indicated by circles. (d) Wave functions of non-resonant states with energies positions indicated by circles in (c).}
\label{fig_3_1}
\end{figure}

The idea of using reflection coefficients to describe the leaking rate (or tunneling rate) of carriers in the active region is rather simple. In the TMM model, an incident wave from the contact serves as the input of the calculation. Normally, we assign the amplitude of the incident wave to be one at a certain available channel which is determined from the complex-band structure\cite{COMX} (including real $k$ solution for the propagating waves along the carrier-transport direction and complex $k$ for evanescent waves). In this manner, the amplitude reflected and transmitted can be easily calculated by comparing with the amplitude of the assigned incident wave. If the system is composed of non-absorbing material, the sum of the transmission and reflection coefficient is one. If there is a finite absorption in the system, the incident wave is absorbed by the material with an amount proportional to the probability of the electron being inside the bounding area. In this work, we artificially introduce a small damping in the system, and a sharp dip appears in the reflection spectra due to the fact that the energy of incident wave matches that of a bound or quasi-bound state. The incident wave can be viewed as a probe, which excites a bound state when their energies match, and that leads to a strong absorption due to the relatively large magnitude of the bound state in the active region. As for the incident waves with energies which do not match any bound states or quasi-bound states, these incident waves bounce back with almost full magnitude and contribute relatively large reflection coefficients in the reflectance spectra. This method allows us to localize the energies of bound states (or quasi-bound state, depending upon the ratio of the wave function amplitude across the interface between the barrier and active region). At the same time, the actual wavefunction can be generated from the TMM. Note that the artificial damping parameter, $\gamma$ introduces energy broadening in the energy spectra. The value of $\gamma$ is inversely proportional to the carrier lifetime. Therefore, the $\gamma$ cannot be larger than the actual energy broadening observed in the experimental spectra. For illustration, we show the calculated reflection spectra of an InN/InGaN multi-QW structure for both the quasi-bound states and the non-resonance states  in Fig. (\ref{fig_3_1}). Fig. (\ref{fig_3_1}a,b) shows the case when the selected energies are in resonance with quasi-bound states, while  Fig. (\ref{fig_3_1}c,d) shows the case when the selected energies are out of resonance. The relative strength of leaking rates are assumed proportional to the depths of the dips.

\section{Determination of Band Bending and Dominant Photon Absorption Process}

\begin{figure}[H]
		\centering\includegraphics[width=1\columnwidth]{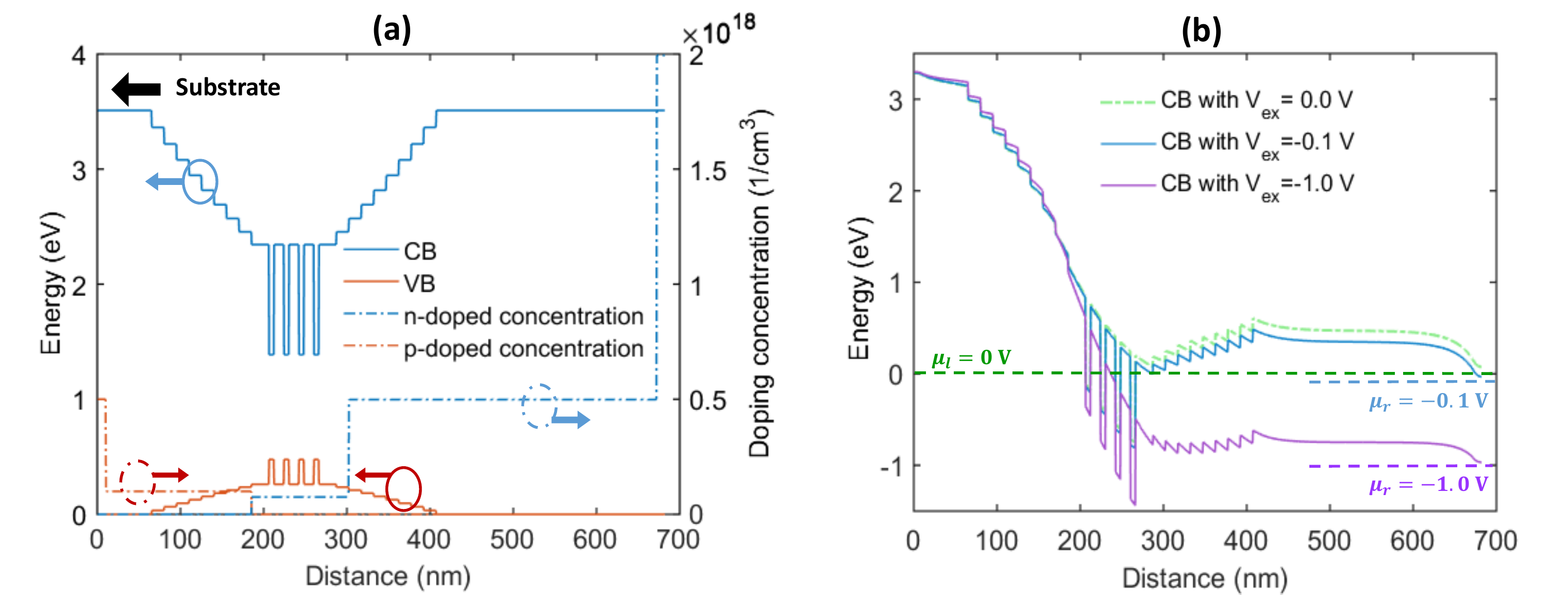}
	\caption{(a) Flat-band diagram for InGaN photodiode. (b) Self-consistent band diagram under various external bias calculated by NEGF.}
\label{fig_5_1}
\end{figure}

In this section, we perform the TMM simulation, which is described in Sec. (5),  to calculate the leaking rate spectra considering the full InN/InGaN disk-in-wire photodiode and determine the dominant photon absorption process for the photocurrent. Given the indium mole fraction in the disk region, strain-induced deformation potential, and the polarization field obtained from single disk-in-wire calculation, the full-device structure can be built up by repeating the potential profile of the single disk-in-wire structure four times to account for four disks in one nanowire structure, and sandwiching it between graded-index p- and n-doped layers according to the epitaxial structure in Ref. \citenum{INTRO_11}. To incorporate the electric field contributed from both the external bias and the built-in potential of the p-i-n structure, we use the self-consistent transport model based on non-equilibrium Green's function (NEGF) to estimate the band-bending effect induced from both carrier injection and doping concentration. Figure (\ref{fig_5_1}) shows both the flat-band diagram (Fig. (\ref{fig_5_1}.a)) and the band diagram under non-equilibrium conditions (Fig. (\ref{fig_5_1}.b)). Due to the unintentionally n-doping, the central intrinsic region is assumed to have a n-type concentration of $7.5\times 10^{-16}$ cm$^{-3}$. This unintentionally n-doped concentration is of importance since it influences not only the band diagram in the active region, but also the relative band bending between the active region and contact, that further affects the tunneling behavior of carriers.

\begin{figure}[H]
		\centering\includegraphics[width=1.\columnwidth]{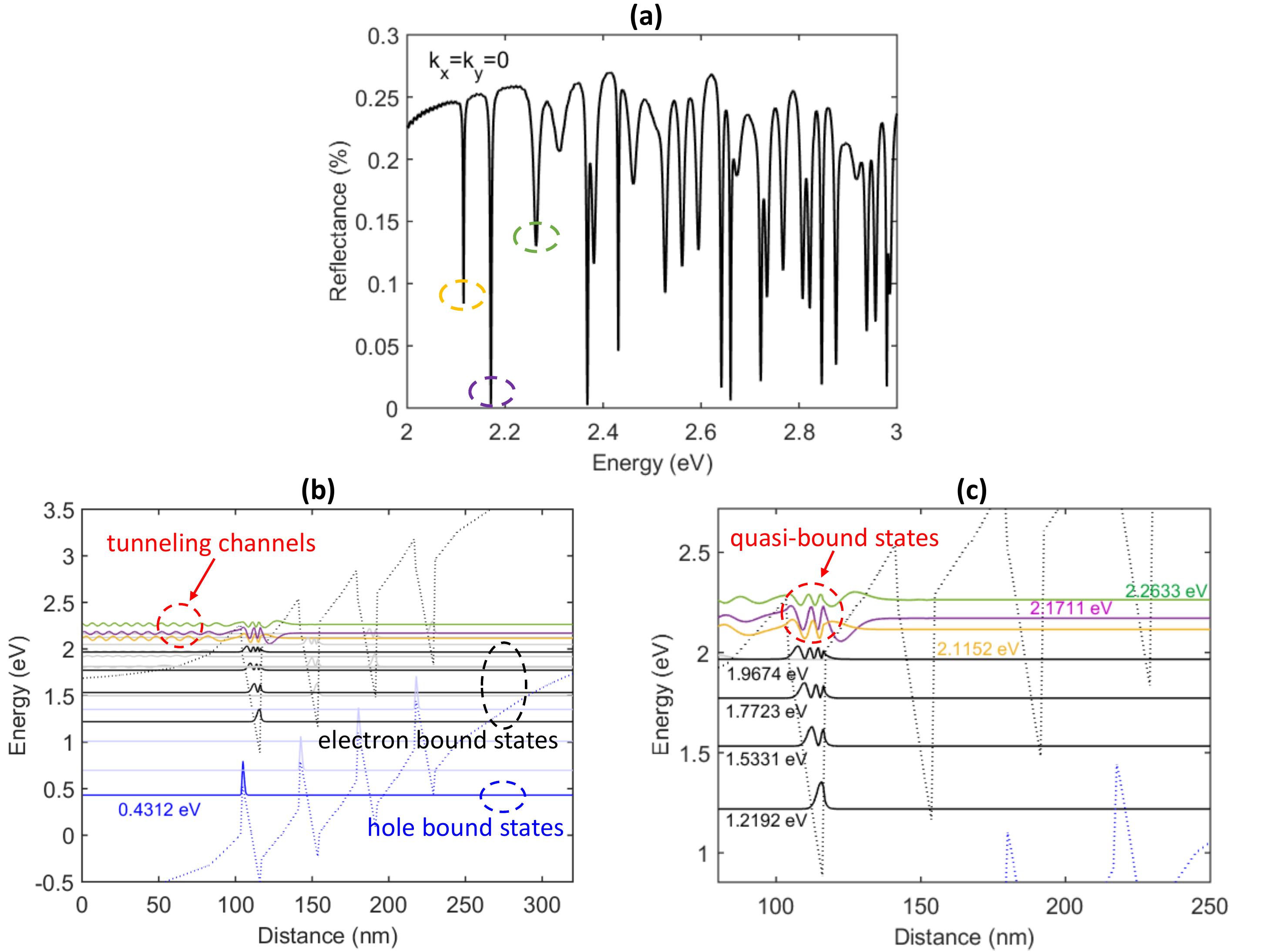}
	\caption{(a) Reflection spectra in the In$_{0.91}$Ga$_{0.08}$N/In$_{0.4}$Ga$_{0.6}$N multi-QW structure for =0 obtained by TMM. (b) Low-lying quasi-bound and bound states in the multi-QW structure. (c) Enlarged view for the quasi-bound states and bound states in multi-QW region.}
\label{fig_5_2}
\end{figure}

Given the self-consistent potential under non-equilibrium condition including the spatially dependent doping profile,the leaking-rate spectra can be calculated using EBOM-based TMM. We start with the leaking-rate spectra of the In$_{0.91}$Ga$_{0.08}$N/In$_{0.4}$Ga$_{0.6}$N disk-in-wire guided wave photodiode at the zone center ($k_x=k_y=0$), which is shown in Fig. (\ref{fig_5_2}.a). Figures (\ref{fig_5_2}.b) and (\ref{fig_5_2}.c) show the first three quasi-bound states corresponding to the first three dips in the leaking-rate spectra (Fig. (\ref{fig_5_2}.a)), the first hole bound states, as well as the electronic bound states at each QW region.  By examining the energy separations between bound states and quasi-bound states, one can determine the possible photon absorption process in such structure.

There are typically three major mechanisms which contribute to the photocurrent, and one can roughly distinguish them by examining the energy separation and photocurrent peak positions. The first one is the simple one-photon absorption process, in which the electron transits into a quasi-bound state in the conduction band and contributes to photocurrent via the direct tunneling process. In this case, the photocurrent peak positions are determined by the SPA spectra multiplied by tunneling rate of each quasi-bound state. The second mechanism is photoabsorption followed by phonon-assisted tunneling. Namely, carriers in the confined region jump to an excited state via photoabsorption, then escape from the active region by absorbing phonons. In this case, the energy separation between the initial and final states should match the photon energy plus the energy of available phonon in the material, and the photocurrent peak positions are determined by the single-photon absorption (SPA) process involving a bound state and an excited state below the tunneling threshold by a phonon energy. The third mechanism is two-photon absorption (TPA) process. The carrier is initially in the valence band or conduction band, and jump into quasi-bound states by absorbing two photons. In this case, both photons involved in the TPA process have identical energy, and the photocurrent peak position is determined by the single photon energy.

We first look into the SPA process. The minimum possible SPA energy is the transition involving the first quasi-bound level in conduction band and top level in valence band. From Fig (\ref{fig_5_2}.c), the separation between these two states is $2.1152-0.4312=1.6840$ eV, which is much larger than the first measured photocurrent peak, which is around $1$ eV. Even with the phonon-assisted photoabsorption, the threshold for photocurrent can only be lowered by the available phonon energy, which is close to 26 meV at room temperature for the experimental condition considered here.
As a result, the dominant absorption process which contributes to the photocurrent below $1.7$ eV is mainly caused by the TPA, although SPA still contributes to photocurrent at the higher energy region. Thus, both the SPA and TPA need to be considered in this work for calculating the photocurrent spectra generated by In$_{0.91}$Ga$_{0.09}$N disk-in-wire photodiode.

\section{One-Photon and Two-Photon Absorption}
The optical properties of interest include both the inter-band transition and intra-band transition. Here we briefly describe the derivation for the momentum dependent optical matrix elements from the EBOM Hamiltonian. The detailed derivation for the case of ZB material was reported in Ref. \citenum{EBOM2}. We extend the derivation by using the EBOM Hamiltonian suitable for WZ materials\cite{EBOM3}.

The optical matrix elements between EBOM basis ($|s\rangle, |x\rangle, |y\rangle,$ and $|z\rangle$) is given by
\begin{equation}
\frac{\hbar}{m_0}\langle \alpha',\mathbf{k}|\hat{\beta}\cdot\mathbf{p}|\alpha,\mathbf{k}\rangle=i\langle \alpha',\mathbf{k}|\hat{\beta}\cdot[H, \vec{\rho}]|\alpha,\mathbf{k}\rangle=i\langle \alpha',\mathbf{k}|\hat{\beta}\cdot[H_{ebom}^{8\times8}, \vec{\rho}]|\alpha,\mathbf{k}\rangle.
\label{eq_6_1}
\end{equation}
where the unit vector $\hat{\beta}$ represents the polarization of the incident light.
In reciprocal space, we can write the position vector $\vec{\rho}$ as an operator $i\nabla_{\mathbf{k}}$, and hence we have
\begin{equation}
[H^{8\times8}_{ebom},\vec{\rho}]=-i\nabla_{\mathbf{k}}H^{8\times8}_{ebom}(\mathbf{k}),
\label{eq_6_2}
\end{equation}
which leads to a simple analytic expression suitable for $\mathbf{k}$ in the full BZ,
\begin{equation}
P^{\alpha,\alpha'}_{\beta}(\mathbf{k})=\hat{\beta}\cdot\nabla_{\mathbf{k}}H^{8\times8}_{ebom}(\mathbf{k}).
\label{eq_6_3}
\end{equation}
Note that Eq. (\ref{eq_6_3}) can be used for the calculation of both inter-band transition and intra-band transition.

Given the optical matrix elements, we can investigate the optical properties of In$_{x}$Ga$_{1-x}$N QW through the dielectric function $\epsilon\equiv\epsilon_1+i\epsilon_2$. The imaginary part $\epsilon_2$ for the polarization is given by\cite{TPA_2}
\begin{equation}
\begin{aligned}
 \epsilon_2(\omega)&=\frac{2\pi e^2}{V\omega^2}\sum_{k_{||};i,j}|\langle\mathbf{k}_{||},i|P_{\beta}|\mathbf{k}_{||},j\rangle|^2\delta(E_j(\mathbf{k}_{||})-E_i(\mathbf{k}_{||})-\hbar\omega)\\
 &=\frac{e^2}{2\pi d\omega^2}\sum_{i,j}\int d^2k_{||}|\langle\mathbf{k}_{||},i|P_{\beta}|\mathbf{k}_{||},j\rangle|^2\delta(E_j(\mathbf{k}_{||})-E_i(\mathbf{k}_{||})-\hbar\omega)
 \end{aligned},
 \label{eq_6_4}
\end{equation}
where $|\mathbf{k}_{||},i\rangle$ denotes the $i$th electronic state of the quantum well associated with wave vector $\mathbf{k}_{||}$, $\hbar\omega$ denotes the photon energy, and $V$ denotes the volume of the slab, which can be written as the cross-sectional area multiplied by the slab thickness. Due to the translational invariance in the lateral direction $(x,y)$, the summation over $\mathbf{k}_{||}$ is replace by integral $A/4\pi^2\int d^2 k_{||}$. The real part of the dielectric function ($\epsilon_1$) can be obtained from the imaginary part ($\epsilon_2$) by utilizing the Kramers-Kronig relation. The one-photon absorption coefficient is related to $\epsilon_2 (\hbar\omega)$ by
\begin{equation}
	\alpha(\hbar \omega)=\frac{\omega}{cn_r}\epsilon_2(\hbar\omega),
	\label{eq_6_5}
\end{equation}
where $c$ is the speed of light, $n_r$ is the refractive index of the active material.

The TPA coefficient ($\beta_2$) for the QW is given by \cite{TPA_1}
\begin{equation}
    \begin{aligned}
	\beta_2=\frac{8\pi\omega}{c^2\epsilon_1(\omega)V}\sum_{\mathbf{k}_{||},mn}|M_{mn}(\mathbf{k}_{||})|^2
	        \times\delta[2\hbar \omega-E_n(\mathbf{k}_{||})+E_m(\mathbf{k}_{||})],
	\end{aligned}
	\label{eq_6_6}
\end{equation}
where
\begin{equation}
\begin{aligned}
	& M_{mn}(\mathbf{k}_{||})=
	& 2(\frac{e}{m_e\omega})^2\sum_{s}\frac{\langle n \mathbf{k}_{||}|\hat{\epsilon}\cdot\mathbf{p}|s\mathbf{k}_{||}\rangle \langle s \mathbf{k}_{||}|\hat{\epsilon}\cdot\mathbf{p}|m\mathbf{k}_{||}\rangle}{\hbar\omega-[E_s(\mathbf{k}_{||})-E_m(\mathbf{k}_{||})]+i\Gamma}.
\end{aligned}
	\label{eq_6_7}
\end{equation}
$E_m(\mathbf{k}_{||})$, $E_s(\mathbf{k}_{||})$, and $E_n(\mathbf{k}_{||})$ denote the energies of initial, intermediate, and final state, respectively. The intermediate states, $| s\mathbf{k}_{||} \rangle$, can be either conduction or valence subband states. For all transitions, the initial state must be occupied and the final state must be empty. For simplicity, we have not included the effect of thermal excitation, although it is not negligible at room temperature. Such an effect will be left for future research. To avoid the singularity coming from the denominator of the second-order perturbation theory, an imaginary number $(i\Gamma)$ is introduced in Eq. (\ref{eq_6_7}). The linewidth parameter $\Gamma$ is related to the finite lifetime of the QW states. All the optical constants for III-N materials used in this work are adopted from Ref \citenum{eps_1}.

In order to capture all possible transitions in the device, all the states corresponding to different $\mathbf{k}_{||}$'s are needed in the calculation of absorption. For WZ materials, both the band structure and optical matrix elements have strong anisotropy. As a result, the $\mathbf{k}_{||}$  integration cannot be avoided in this case. There are several ways to conduct the $\mathbf{k}_{||}$ integration, including the method of selecting a special ${\bf k}$ point with proper weighting according to the symmetry of crystal \cite{BZ_CUT_1}, or simply doing a dense-mesh zone integration in the desired portion of first BZ. In this work, we apply the zone-integration method with area-dependent weighting to capture the full transition characteristics in the first  BZ. Due to the $C_{3}$ symmetry of WZ crystal, we only sample  $\mathbf{k}_{||}$ within the triangle with vertices at $\Gamma$, $K$, and $M$ points as shown in Fig (\ref{fig_6_1}).

The other effect introduced by the lateral confinement is the energy quantization in the transverse direction. In the disk-in-wire system, both the interface defined by the high-indium QD and the nanowire structure impact confinement. For the InGaN QD interface, the quantum confinement, the deformation potential induced by strain, piezoelectric field, and the material band offset have been taken into account in our model calculation. Again, due to the fairly large wire structure in our system (radius $=~26 nm$), the confinement effect is expected to be small, and the distribution of available $\mathbf{k}_{||}$ states in reciprocal space can be approximated by a  continuum. However, in the low energy region, where the spacing between available $\mathbf{k}_{||}$ states are relatively large, and the energy quantization still makes a noticeable modification to the absorption spectra.

\begin{figure}[H]
		\centering\includegraphics[width=1\columnwidth]{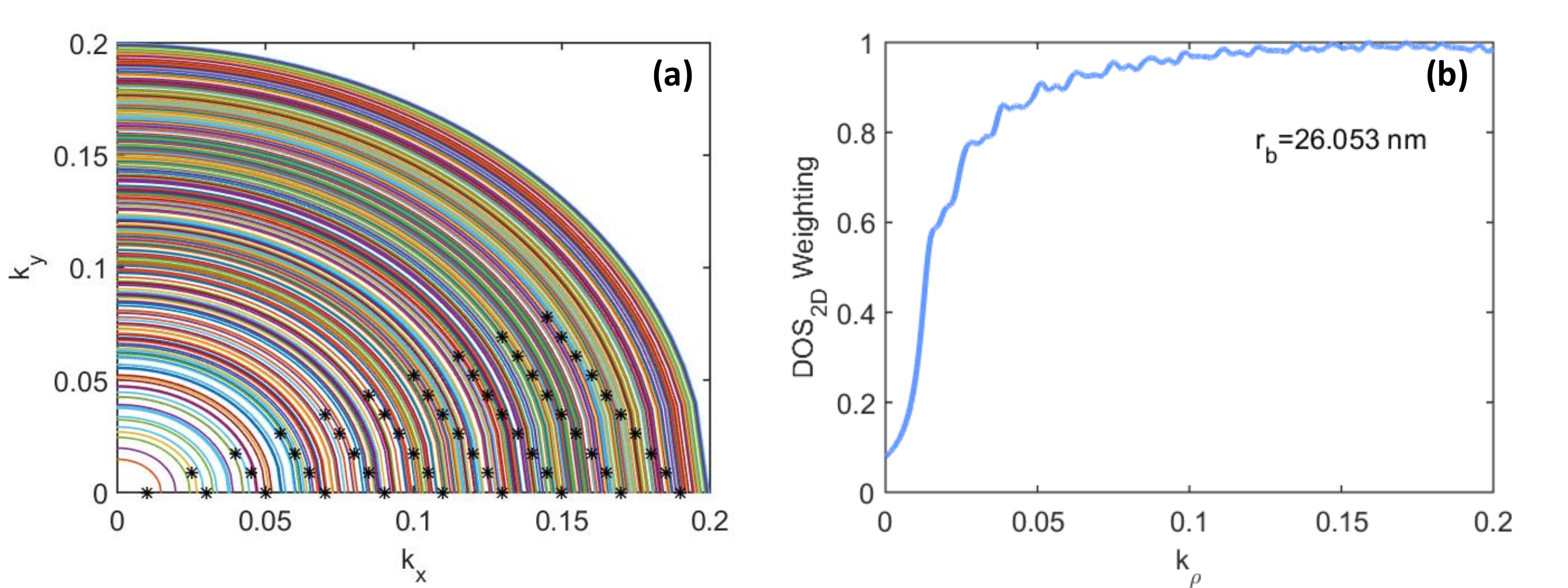}
	\caption{(a) Quantized states introduced by a cylindrical hard-wall boundary condition due to lateral confinement. (b) Density-of-states (broadened by a finite width) of laterally confined states for a cylinder of radius $r_0=26.053$ nm with infinitely high barrier.}
\label{fig_6_2}
\end{figure}

In Fig. (\ref{fig_6_2}.a), we show the distribution of allowed $\mathbf{k}_{||}$'s  which satisfy the boundary condition $J_n(kr_0)=0$ for a nanowire structure with radius $r_0=26.053$ nm based on the truncated-crystal approximation. In this work, we first generate a DOS curve defined by $D(k_{\rho})=\sum_{\mathbf{k}_{||}}\delta(k_{\rho}-k_{||})$ as a function of the magnitude of $k_{\rho}$ by introducing a broadening in the delta functions (Fig. (\ref{fig_6_2}.b)). We then use interpolation to obtain the weighting factor that accounts for the DOS due to lateral confinement at each sampling point of $\mathbf{k}_{||}$ as shown by the black dots in Fig. (\ref{fig_6_1}.b). Due to large spacings among allowed states in the low energy region, the weighting factor that reflects the correction due to lateral confinement is relatively strong near the zone center. To summarize, we have introduced two weighting factors: one accounts for the zone integration effect and the other accounts for the energy squeezing effect induced by lateral confinement.

\section{Photon Density-of-States in a Graded-Index Waveguide}

The graded-index layers of the epitaxial design in the considered InN/InGaN disk-in-wire photodiode only allows photons with certain energies to propagate in the layer of nanowire arrays. The peaks of the photocurrent spectra occur when the photon is in resonance with  cavity modes, and the absorption strength at those resonance energies is significant, which indicates good electron-photon coupling in the active region. As a result, the DOS of optical waves propagating in the photodiode can also influence the peak positions in the photocurrent spectra.

\begin{figure}[H]
		\centering\includegraphics[width=1\columnwidth]{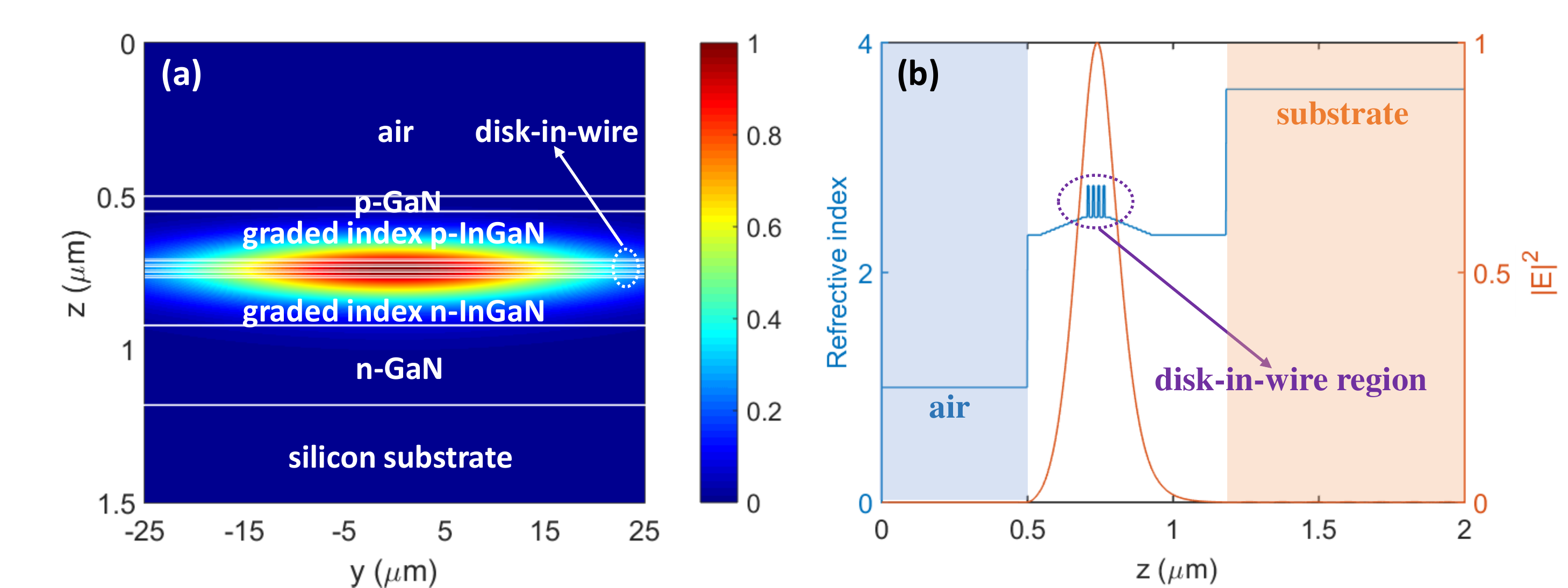}
	\caption{(a) Schematic diagram and the fundamental model profile of the InGaN disk-in-wire waveguide structure considered in the RCWA simulation. (b) Schematic diagram of spatial-dependent refractive index of the InGaN disk-in-wire waveguide structure.}
\label{fig_7_0}
\end{figure}

\begin{figure}[H]
		\centering\includegraphics[width=0.8\columnwidth]{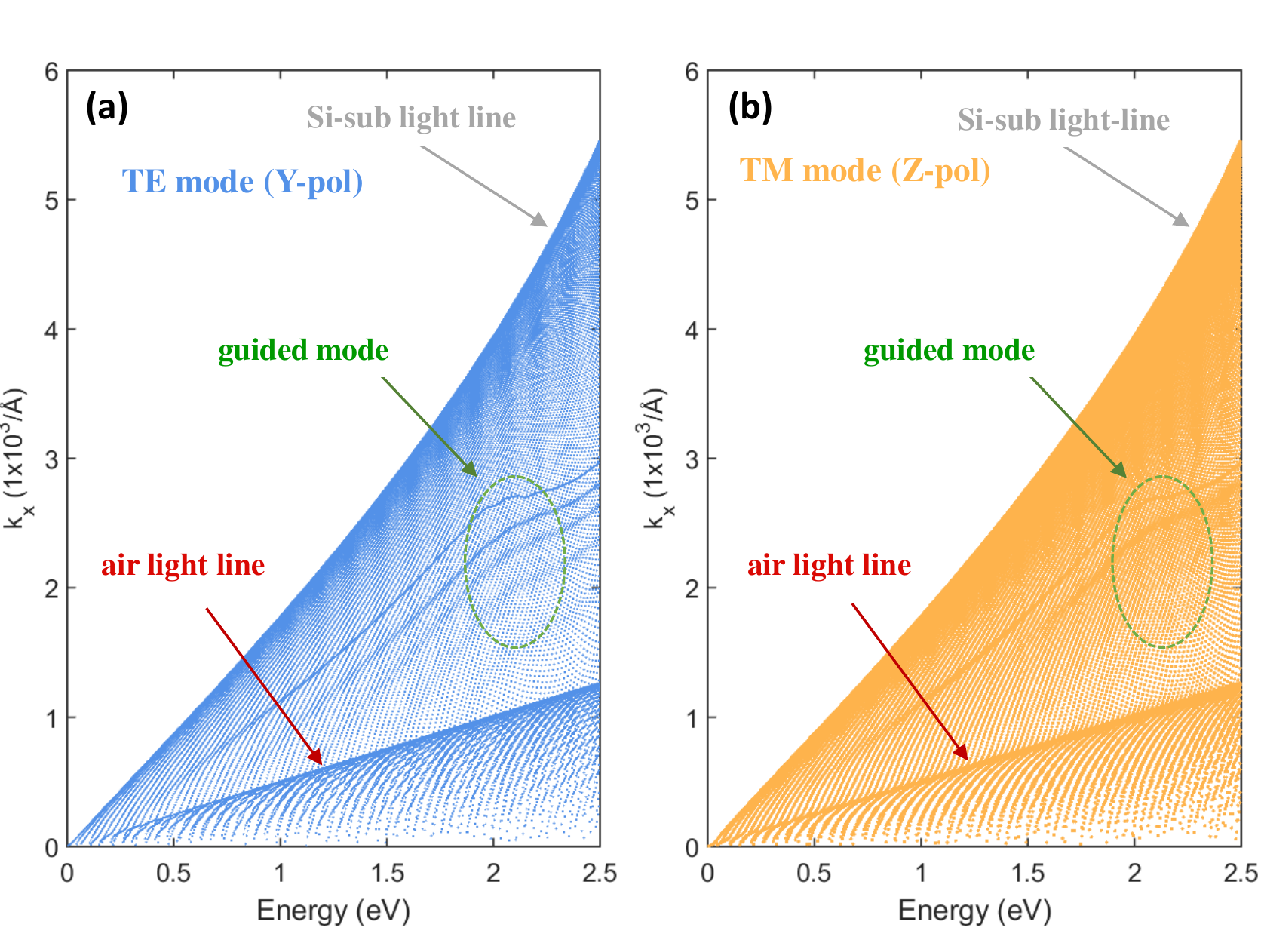}
	\caption{Photonic band structure of the In$_{0.91}$Ga$_{0.09}$N/In$_{0.4}$Ga$_{0.6}$N graded-index waveguide structure for (a) TE mode component and (b) TM mode component.}
\label{fig_7_1}
\end{figure}

The considered graded-index waveguide structure and the corresponding refractive index of each layer at $\sim 1$ eV are shown in Fig. (\ref{fig_7_0}.a) and Fig. (\ref{fig_7_0}.b), respectively. It consists of n-/p-doped In$_x$Ga$_{1-x}$N graded-index layer with $x=0.4 \sim 0.04$, and disk-in-wire array at the center as the active layer. The entire graded-index structure is sandwiched by p-/n-doped GaN cladding layers. A thick silicon substrate layer and an air layer are also incorporated in our simulation in order to capture the accurate dispersion curves for both the guided mode and resonance mode. In our simulation, a three-dimensional structure consisting of two-dimensional disk-in-wire array with wire width $= 60$ nm and fill factor of $0.91$ is considered. The photonic band structure of the In$_{0.91}$Ga$_{0.09}$N/In$_{0.4}$Ga$_{0.6}$N disk-in-wire photodiode inside a graded-index waveguide structure is calculated based on the RCWA method. Since the dielectric constants of materials involved are functions of the photon energy, the input photon energy is scanned for the solution of $k_x$ that satisfies the Maxwell's Equations. The results are shown in Fig (\ref{fig_7_1}) for both TE ($y$-polarization) and TM ($z$-polarization) {modes}. Each point in Fig. (\ref{fig_7_1}) corresponds to a {propagating} mode in the In$_{0.91}$Ga$_{0.09}$N/In$_{0.4}$Ga$_{0.6}$N disk-in-wire waveguide with a certain propagation constant {($k_x$)}. The shadow in the background corresponds to the intrinsic modes in constituent bulk materials {including the} silicon substrate and air (slab mode). The heavy dark curves {describe} the dispersion of the guided wave associated with the graded-index structure. The dispersion curves have a finite width since they can still leak into the substrate.  The width of the dispersion curves in the high-energy region becomes narrower as they represent the well-confined modes in the device. In regions where the dispersion curves are broadened, the photon modes are more leaky, which are identified as resonance modes. If the silicon substrate is removed in the simulation, the typical guided mode dispersion curves above the air light line {become clear sharp lines}, while the continuum modes, which correspond to the radiation modes, {appear below the air light} line.

\begin{figure}[H]
		\centering\includegraphics[width=0.8\columnwidth]{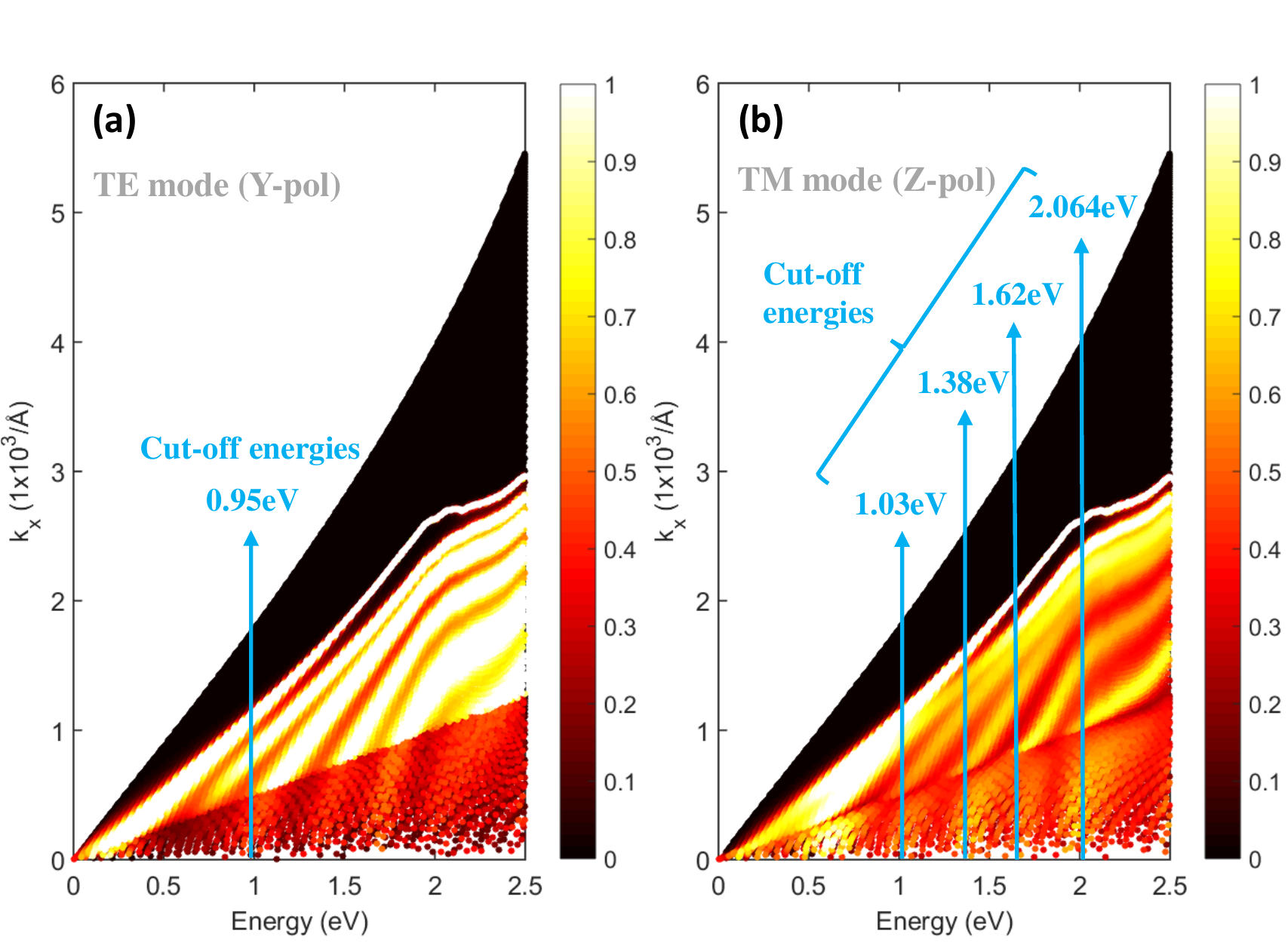}
	\caption{The normalized modal density in the nanowire layer of the In$_{0.91}$Ga$_{0.09}$N/In$_{0.4}$Ga$_{0.6}$N graded-index waveguide structure for (a) TE and (b) TM polarization. The color bar indicates the probability strength of a mode in the nanowire layer}
\label{fig_7_2}
\end{figure}

It is noted that for confined modes the photon with small propagation constant (below air light line) tend to have relatively large DOS due to the small slope ($dE/dk_x$) of the dispersion curve near zone-center, especially for TM modes\cite{TM_DOS}. As a result, the optical transition process are more involving with the less-confined modes near $k_x=0$ than the well-confined modes above the air light line. In our case, since the fill factor of the disk-in-wire array is fairly large ($\sim 0.91$), the modal dispersion curves below the air light line are {difficult to distinguish from the slab modes}. However, confined modes in this region with small group velocity along the $x$-axis still exist in the {present} structure. To help sort out the characteristics of the resonance modes, the probability distribution of each mode is calculated in each region of the device structure and show their modal density in the nanowire layer for TE and TM modes in Fig. (\ref{fig_7_2}). By comparing Fig. (\ref{fig_7_2}.a) and (\ref{fig_7_2}.b), the TM modes show over-all stronger modal density in the nanowire array region. As a result, we expect the effect of the photon DOS is mainly contributed by TM modes, and only TE modes with cut-off energy near $0.95$ eV and $1.6$ eV have comparable modal density with those of TM modes. However, since the absorption above $\sim 1$ eV is dominated by TM modes (which will be discussed in the next section), we only include the TE-mode contributions  with cut-off energies near $1$ eV in the photocurrent calculation. Based on the discussion above, we shall quantitatively characterize the effect of the photon DOS by the segments of dispersion curves below the air light line. The photon DOS for each resonance mode are fitted with the Lorentzian functions with the center of functions corresponding to the cut-off energy and the broadening width determined by the spread of each group of dispersion curves near $k_x=0$. From Fig. (\ref{fig_7_2}.b), we determine the cut-off energies of the TM modes that are of interest as $1.03$ eV, $1.38$ eV, $1.62$ eV, and $2.06$ eV, while the corresponding broadening widths are $0.4$ eV, $0.9$ eV, $0.4$ eV, and $1.7$ eV, respectively. The TE mode with {$0.95$ eV cut-off energy} and $=0.4$ eV broadening width is also included (Fig. (\ref{fig_7_2}.a)). The fundamental mode and first excited mode for both TE and TM polarizations are not selected due to their poor modal density with TPA absorption spectra. Note that both TE and TM resonance modes with cut-off energy near $\sim 1$ eV exist in the photonic band structure of the waveguide. As a result, the DOS near $1$ eV, which consists of two Lorentzian functions attributed to TE and TM components, is about twice as strong compared to others in this work.
\section{Results and Discussions}
\begin{figure}[h!]
		\centering\includegraphics[width=1\columnwidth]{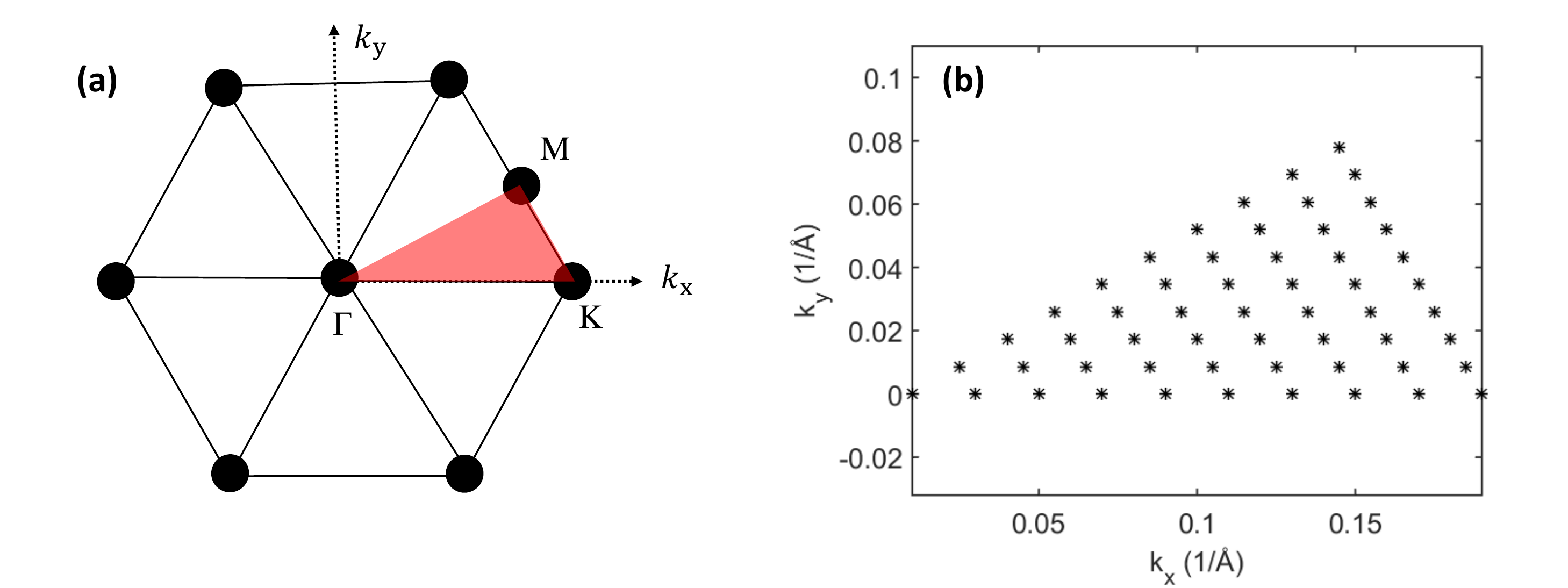}
	\caption{(a) Brillouin zone of a WZ quantum well. The grey area indicates the irreducible segment. (b) Sampling points of $\mathbf{k}_{||}$ in the irreducible segment used in the integration over $\mathbf{k}_{||}$.}
\label{fig_6_1}
\end{figure}

\begin{figure}[H]
		\centering\includegraphics[width=1\columnwidth]{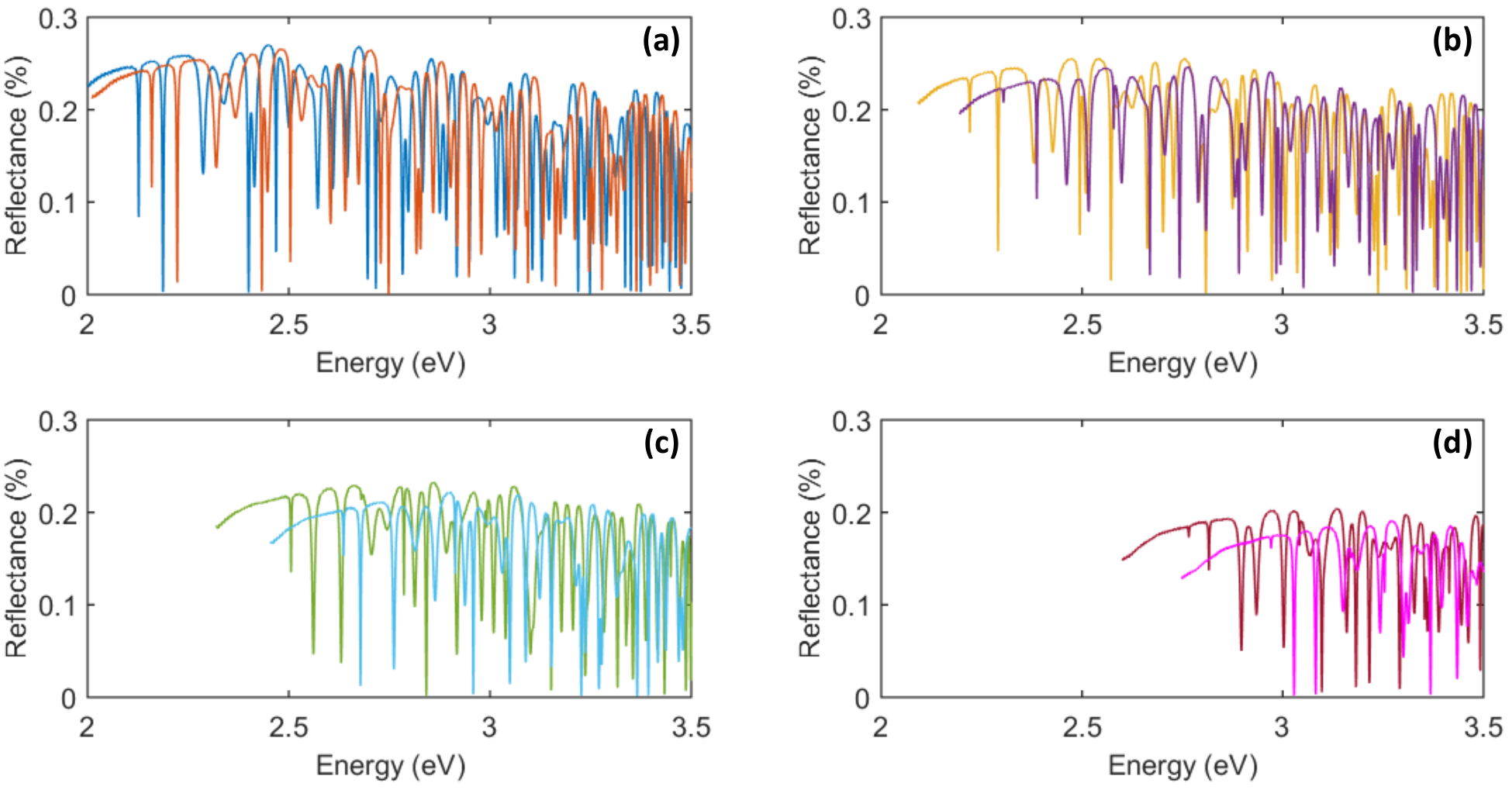}
	\caption{Individual contributions to the reflection spectra from various sampling points for $\mathbf{k}_{||}$, calculated by TMM with EBOM. (a)$\mathbf{k}_{||}=(0.0177,0.0)$, $(0.0443,0.0154)$. (b)$\mathbf{k}_{||}=(0.071,0.031)$, $(0.0975,0.046)$. (c)$\mathbf{k}_{||}=(0.124,0.061)$, $(0.151,0.077)$. (d)$\mathbf{k}_{||}=(0.177,0.092)$, $(0.204,0.107)$.}
\label{fig_8_1}
\end{figure}

In this section, we discuss how to combine the carrier tunneling rate with the absorption spectra to get the  photocurrent spectra. A selection of $55$ sampling points are chosen for the wave vector $\mathbf{k}_{||}$  within  $1/12$ of the first BZ as illustrated by dark dots in Fig. (\ref{fig_6_1}.b). The reflection spectra is then calculated at each sampling $\mathbf{k}_{||}$ by using EBOM-based TMM described in Sec. (5). In Fig. (\ref{fig_8_1}), we show the reflection spectra with the damping parameter $\gamma=10$ meV for various sampling wave vectors. Each dip in the reflection spectra reflects significant absorption at that energy, indicating the existence of quasi-bound or bound state. The energies and wavefunctions corresponding to these dips are then collected and a set of quasi-bound states is constructed, which serve as available tunneling channels for carriers to reach the contact. The relative strengths of leaking rates are assumed proportional to the depths of the dips. As for the calculation of absorption spectra, since the experiment was conducted with the light incident from the $y-z$ facet, only the absorption corresponding to $y$-polarized and $z$-polarized photons is considered. By multiplying the leaking rate of a tunneling state and the absorption coefficient for a transition into that state, the contribution from that transition to the photocurrent spectrum is obtained. Integrating all these transitions over the sampling points in $\mathbf{k}_{||}$ gives rise to the total photocurrent spectrum.

\begin{figure}[H]
		\centering\includegraphics[width=1.0\columnwidth]{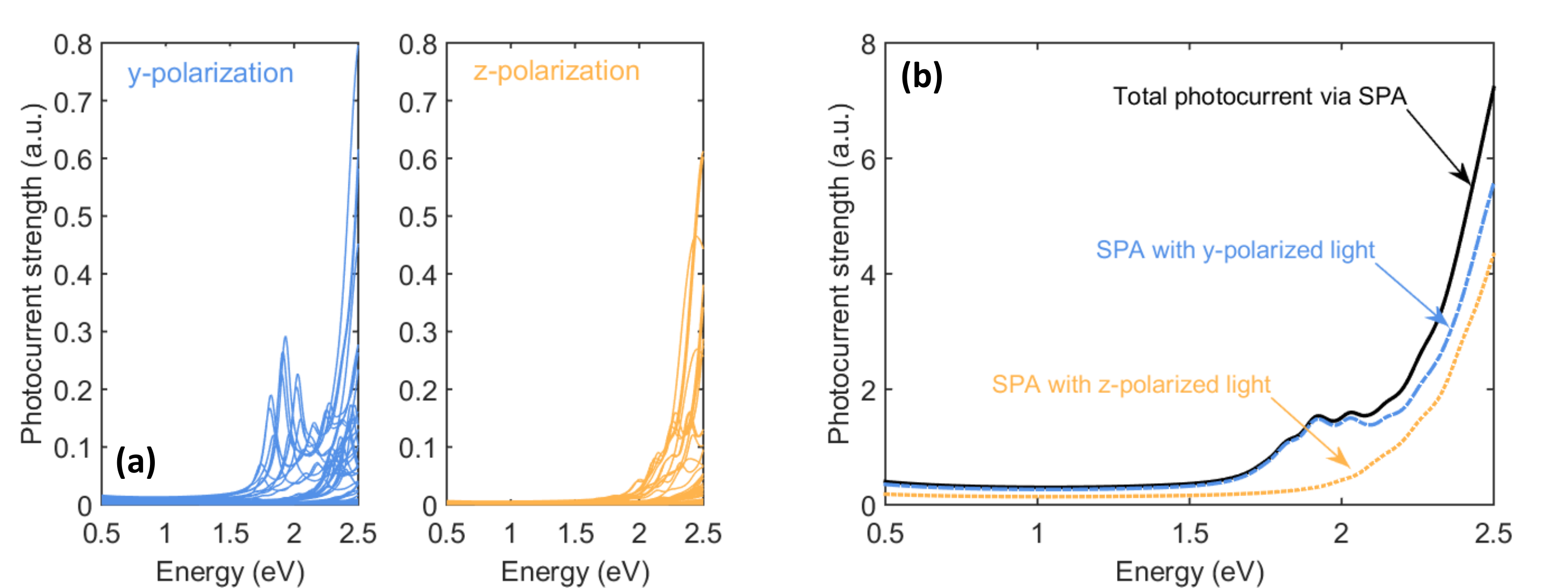}
	\caption{Photocurrent spectra via SPA process with both  y-polarized and z-polarized light. (a) Individual contributions from various sampling points for $\mathbf{k}_{||}$. (b) Total SPA photocurrent spectra.}
\label{fig_8_2}
\end{figure}

Figure (\ref{fig_8_2}) shows the photocurrent spectrum due to SPA for the In$_{0.91}$Ga$_{0.09}$N disk-in-wire structure. The whole spectrum is broadened with $\sigma=0.1$ eV due to the inhomogeneity of the structure. As expected, the first absorption peak appears in the energy region higher than $1.5$ eV. This suggests that the measured photocurrent peaks in low-energy region are attributed to some mechanisms other than SPA. The SPA absorption mainly contributes to the  photocurrent around $2$ eV, where the dominant SPA process is due to the $y$-polarized photon.

\begin{figure}[H]
		\centering\includegraphics[width=1.0\columnwidth]{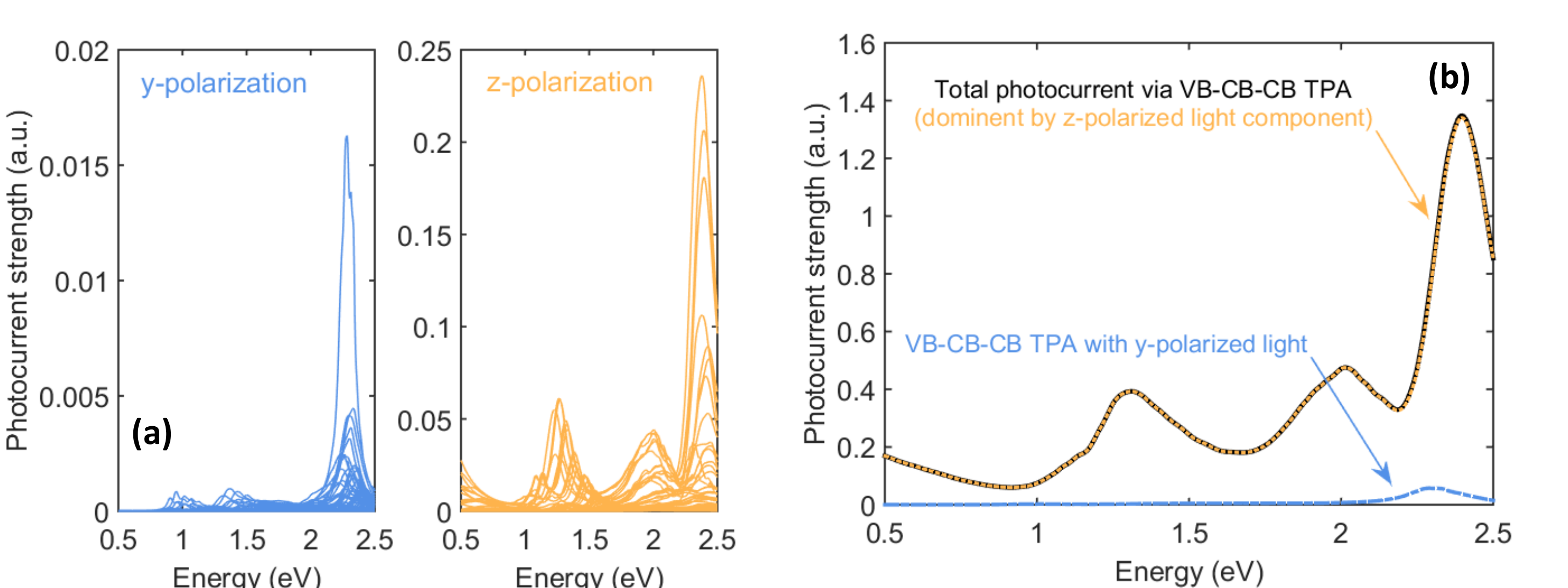}
	\caption{Photocurrent spectra via VB-CB-CB TPA process with both y-polarized and z-polarized light. (a) Individual contributions from various sampling points for $\mathbf{k}_{||}$. (b) Total TPA photocurrent spectra from VB-CB-CB process.}
\label{fig_8_3}
\end{figure}

\begin{figure}[H]
		\centering\includegraphics[width=1.0\columnwidth]{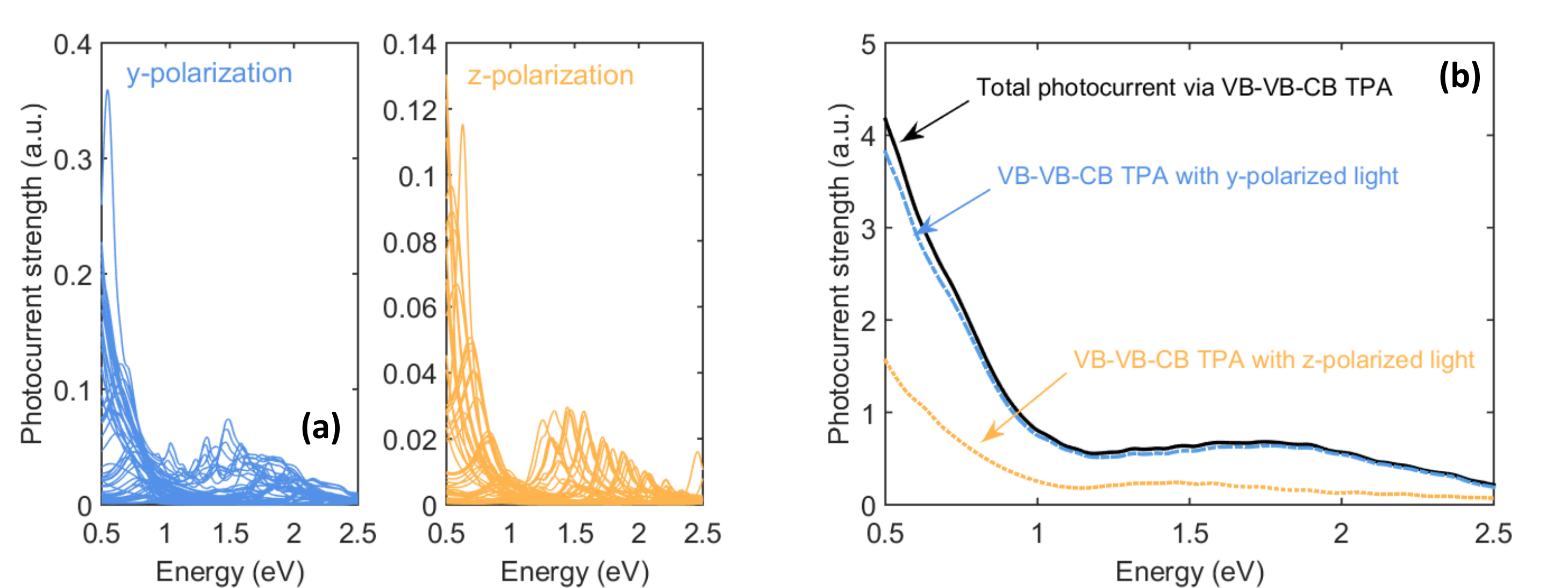}
	\caption{Photocurrent spectra via VB-VB-CB TPA process with  y-polarized and z-polarized light. (a) Individual contributions from various sampling points for $\mathbf{k}_{||}$. (b) Total TPA photocurrent spectra from VB-VB-CB process.}
\label{fig_8_4}
\end{figure}

The TPA processes can be categorized into two different types depending upon the selection of the intermediate state: VB-CB-CB, which uses  conduction subband states as intermediate states, while VB-VB-CB uses valence subband states as intermediate states. Figure (\ref{fig_8_3}) shows the photocurrent spectra via VB-CB-CB absorption of the In$_{0.91}$Ga$_{0.09}$N disk-in-wire structure.  The calculated spectra show a three-packet feature for both $y$- and $z$-polarized photons with $z$-polarization about $10$ times stronger than the $y$-polarization. Figure (\ref{fig_8_4}) shows the photocurrent spectra induced by VB-VB-CB type transition. It is seen that VB-VB-CB  is generally stronger than VB-CB-CB in the low-energy region due to a higher DOS for the valence subbands. The photocurrent strengths for $y$-polarization and $z$-polarization are closer for VB-VB-CB transitions with the $y$-polarization strength about $2.5$ times stronger. Note that the first and second packets due to VB-CB-CB transitions and the packet below $1$ eV due to VB-VB-CB transitions do not really satisfy the double-resonance condition of TPA, but they are enhanced significantly by the $\omega^{-3}$ factor in Eq. (\ref{eq_6_4}). This implies that the photocurrent peaks in the low energy region are pinned and less sensitive to external bias. From Figs. (\ref{fig_8_3}) and (\ref{fig_8_4}), we expect that the photocurrent spectra in the region between $1.2$ eV and $2.5$ eV are mainly contributed by the VB-CB-CB transitions, while VB-VB-CB transitions are dominant in the low energy region ($< 1$ eV).

\begin{figure}[H]
		\centering\includegraphics[width=1.0\columnwidth]{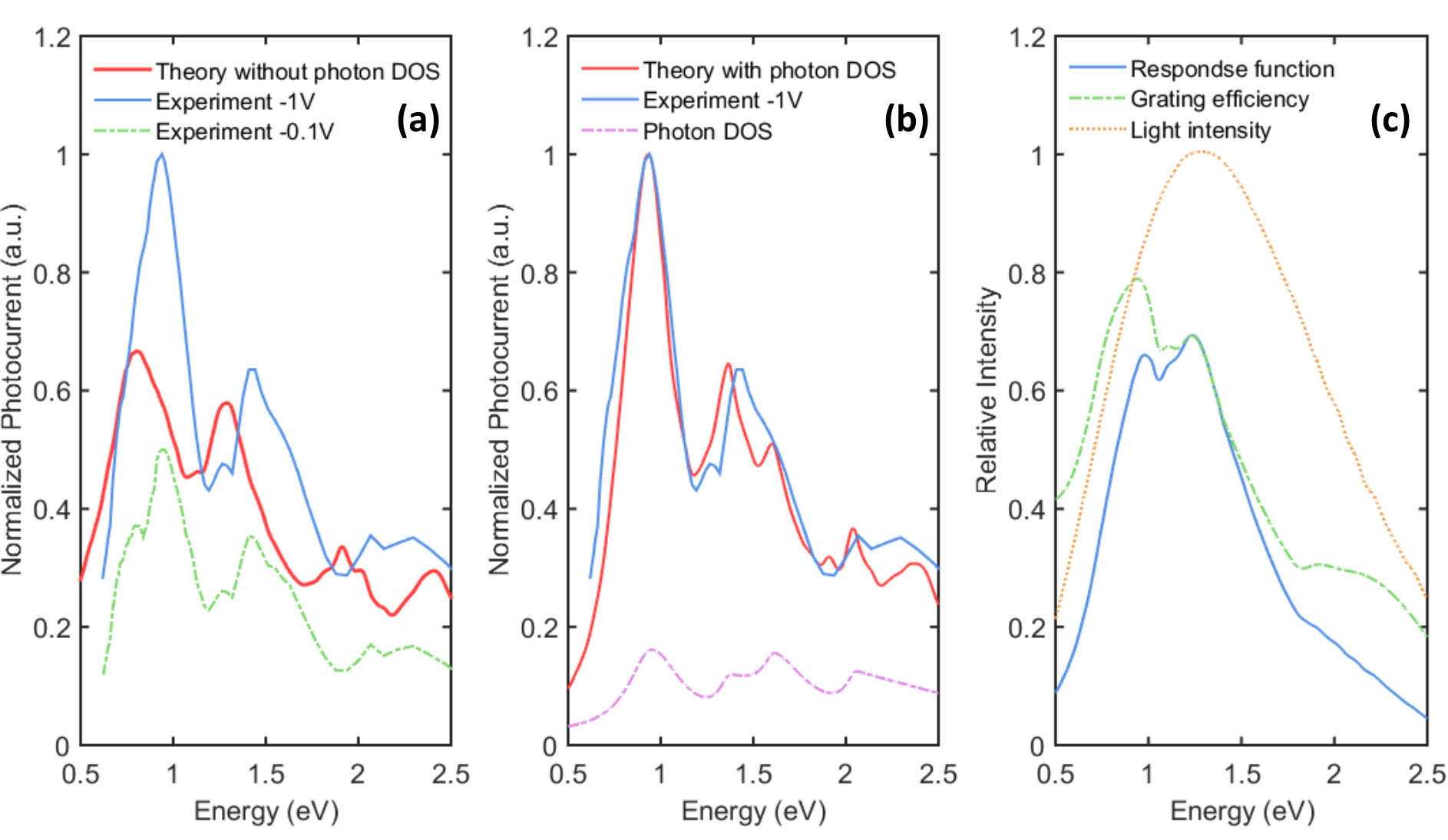}
	\caption{Calculated photocurrent spectra and the comparison with experimental data (a) without including effects due to photon DOS and (b) including effects due to photon DOS. (c) The equipment response functions and light intensity profile used in this work.}
\label{fig_8_5}
\end{figure}

Figures (\ref{fig_8_5}.a) and (\ref{fig_8_5}.b) show the calculated total photocurrent spectra (red line) at $-1$ V bias with and without considering the photon DOS, respectively, as well as their comparison with the experimental data (blue line). Note that in order to compare the calculated photocurrent spectra with experimental data, the energy dependence of the intensity distribution of the light source, the response of detector, and the grating efficiency must be included in the calculation. Those functions adopted from equipment specifications are shown in Fig. (\ref{fig_8_5}.c), and they are incorporated in the calculation. In the case without considering the effect of  the photon DOS (Fig (\ref{fig_8_5}.a)), the peak positions in the calculated spectra match the weak features observed experimentally. The calculated photocurrent spectra shows peaks at $0.806$ eV, $1.287$ eV, $1.914$ eV, $2.009$ eV, and $2.405$ eV. The first peak corresponds to TPA with VB-VB-CB transition. The second and last peak correspond to TPA with VB-VB-CB transitions, while peaks around $2$ eV are induced by SPA. In the measured photocurrent spectra under the $-0.1$ V bias, there are two peaks near $0.8$ eV and $1$ eV, while the one near $0.8$ eV is less noticeable under $-1$ V external bias due to relatively large broadening.

After including the effect of the photon DOS, optical transitions with energies matching the photon DOS peaks in the cavity are greatly enhanced, leading to additional peaks, which can mask the original peaks. As a result, the three packets of peaks in the calculated photocurrent spectrum appear blue shifted, and they are in much better agreement with experimental results at $-1$ V bias. The first peak is shifted to $0.95$ eV and the peak originally at $0.806$ eV in the case without photon DOS correction becomes a small shoulder of the first packet which is also seen in the measured data with external bias $=-0.1$ V. The peak originally at $1.287$ eV in the case without photon DOS correction becomes a shoulder of the second packet, while another peak appears at $1.38$ eV. Besides that, a peak appears at $1.62$ eV after considering the photon DOS. However, due to the small absorption strength in the range from $1.5$ eV to $1.8$ eV, this peak merges into the second packet which is consistent with the experimental data. The second last peak at $2.039$ eV corresponds to VB-CB-CB transition peak at $2.066$ eV coupled with z-polarized photon. The very last peak at $2.375$ eV corresponds to the non-coupling peak at $2.296$ eV with  VB-CB-CB TPA transition in the measured photocurrent spectra. A peak corresponding to SPA is also present at $1.91$ eV in the calculated spectra. We note that the deviations of the lineshape between the calculated spectrum and the measured data between $1.2$ eV and $1.9$ eV are attributed to our approximation of the photon DOS by using a few Lorentzian functions. It may be improved by incorporating the actual photon DOS calculated from the dispersion curve of resonance modes. Furthermore, the ratio of  strengths between SPA and TPA used in the calculation is uncertain, since it depends on light intensity inside the active region, which cannot be determined accurately.

\section{Conclusions}
Comprehensive modeling of the phtocurrent spectrum for an In$_{0.91}$Ga$_{0.09}$N/In$_{0.4}$Ga$_{0.6}$N disk-in-wire guided wave photodiode operating from $1.3$ $\mu$m to $1.55$ $\mu$m has been demonstrated. The model contains strain effects, piezoelectric effects, and the effects of the photon DOS in the calculation. This model has been used to systematically study the electronic, optical, and transport properties in this In$_{0.91}$Ga$_{0.09}$N/In$_{0.4}$Ga$_{0.6}$N disk-in-wire photodiode. Our model successfully predicts photocurrent spectra which agree fairly well with the experimental data. The physical mechanisms for the peaks observed in the experimental photocurrent spectrum are investigated. This is the first direct comparison between the comprehensive theoretical analysis for III-N disk-in-wire structure and the experimental data for the high-indium-content device. The presented approach can be easily applied in III-N materials system with arbitrary complex nanostructures, in which the realistic band structure over the full BZ is incorporated efficiently. Besides that, the presented method utilizing the reflectance spectrum calculation in multi-band TMM method is especially suitable for the simulation of devices driven by optical pumping. It is expected this model can be utilized to study other aspects of III-N devices with ultra-high indium mole fraction. 
\begin{acknowledgement}
This work is supported by the National Science Foundation (MRSEC program), under Grant DMR-1120923, and Ministry of Science and Technology (MOST), Taiwan  under contract no. 109-2112-M-001-046.
\end{acknowledgement}

%\begin{suppinfo}

%\end{suppinfo}


%% graphical table of contents.
%%%%%%%%%%%%%%%%%%%%%%%%%%%%%%%%%%%%%%%%%%%%%%%%%%%%%%%%%%%%%%%%%%%%%

\end{document}